\documentclass{aa}  

\usepackage{graphicx}
\usepackage{subfigure}
\usepackage[breaklinks,colorlinks,citecolor=blue]{hyperref}
\usepackage{txfonts}
\usepackage[flushleft]{threeparttable}

\usepackage{lscape}

\hypersetup{
  bookmarks=true
  citecolor=blue
  colorlinks   = true, %Colours links instead of ugly boxes
  urlcolor     = magenta, %Colour for external hyperlinks
  linkcolor    = blue, %Colour of internal links
  backref = false
  pagebackref = false
}
\usepackage{color}
\usepackage{amstext}

\begin{document}

   \title{Planet signatures and effect of the chemical evolution of the Galactic thin-disk stars}

   \author{Lorenzo Spina
          \inst{\ref{inst1}}
          \and
          Jorge Mel\'endez\inst{\ref{inst1}}
\and
Ivan Ram\'{i}rez\inst{\ref{inst2}}          }

  \institute{Universidade de S\~ao Paulo, IAG, Departamento de Astronomia, Rua do M\~atao 1226, S\~ao Paulo, 05509-900 SP, Brasil - \email{lspina@usp.br}\label{inst1}
  \and Department of Astronomy, University of Texas at Austin; 2515 Speedway, Stop C1400, Austin, TX 78712-1205, USA\label{inst2}
               }

   \date{Received September 22, 2015; accepted October 19, 2015}

% \abstract{}{}{}{}{} 
% 5 {} token are mandatory
 
  \abstract
  % context heading (optional)
  % {} leave it empty if necessary  
   {Studies based on high-precision abundance determinations revealed that chemical patterns of solar twins are characterised by the correlation between the differential abundances relative to the Sun and the condensation temperatures ($T_{c}$) of the elements. It has been suggested that the origin of this relation is related to the chemical evolution of the Galactic disk, but other processes, associated with the presence of planets around stars, might also be involved. }
  % aims heading (mandatory)
   {We analyse HIRES spectra of 14 solar twins and the Sun to provide new insights on the mechanisms that can determine the relation between [X/H] and $T_{c}$.}
  % methods heading (mandatory)
   {Our spectroscopic analysis produced stellar parameters ($T_{eff}$, log~g, [Fe/H], and $\xi$), ages, masses, and abundances of 22 elements (C, O, Na, Mg, Al, Si, S, K, Ca, Sc, Ti, V, Cr, Mn, Fe, Co, Ni, Cu, Zn, Sr, Y, and Ba). We used these determinations to place new constraints on the chemical evolution of the Galactic disk and to verify whether this process alone can explain the different [X/H]-$T_{c}$ slopes observed so far.}
  % results heading (mandatory)
   {We confirm that the [X/Fe] ratios of all the species correlate with age. The slopes of these relations allow us to describe the effect that the chemical evolution of the Galactic disk has on the chemical patterns of the solar twins. After subtracting the chemical evolution effect, we find that the unevolved [X/H]-$T_{c}$ slope values do not depend on the stellar ages anymore. However, the wide diversity among these [X/H]-$T_{c}$ slopes, covering a range of $\pm$4~10$^{-5}$~dex~K$^{-1}$, indicates that processes in addition to the chemical
evolution may affect the [X/H]-$T_{c}$ slopes.}
  % conclusions heading (optional), leave it empty if necessary 
   {The wide range of unevolved [X/H]-$T_{c}$ slope values spanned at all ages by our sample could reflect the wide diversity among exo-planetary systems observed so far and the variety of fates that the matter in circumstellar disks can experience.}


\authorrunning{L. Spina et al.}
\titlerunning{Planet signatures and effect of the chemical evolution of the Galactic thin-disk stars}

   \maketitle
%
%________________________________________________________________

\section{Introduction}
The main motivation of this study has been a recent paper published by \citet{Nissen15} (hereafter, N15), who determined the high-precision abundances of 13 elements in addition to iron in a sample of 21 solar-twin spectra, revealing the existence of a correlation between [X/Fe] and their stellar ages. This work was written in the wake of a sequence of other important studies based on abundance determinations in solar twins and is a significant step forward in study the chemical patterns of solar twins. 

\citet{Melendez09} first showed that the differential chemical abundances of 11 solar twins relative to the solar composition correlate with the condensation temperatures ($T_{c}$) of the elements. They also found that the Sun is depleted in refractory elements with respect to most of the solar twins they analysed.  \citet{Ramirez09} also analysed the chemical content of 22 solar
twins and confirmed the existence of a positive trend in the correlation between [X/H] and the $T_{c}$ (hereafter, [X/H]-$T_{c}$ slope) for 85$\%$ of the stars. Both \citet{Melendez09} and \citet{Ramirez09} speculated that this peculiarity of the solar chemical pattern could be the signature of rocky planet formation that did not occur for the most of the solar twins. They suggested that the positive [X/H]-$T_{c}$ slopes observed in these stars have been produced by accretion onto the star of the refractory material present in the circumstellar disks that is otherwise employed to form planets.

These striking results opened a vivid debate related to the origin of the observed [X/H]-$T_{c}$ slopes. In particular, several teams made great efforts to answer to the following question: is the [X/H]-$T_{c}$ slope connected to the formation of planetary systems around stars, or is it instead related to completely different mechanisms, such as the chemical evolution of the Galactic thin-disk population?  \citet{GonzalezHernandez10,GonzalezHernandez13} claimed that the abundance patterns of solar twins are not affected by the presence of rocky planets around the stars, and \citet{Adibekyan14} clearly revealed a correlation between the steepness of the [X/H]-$T_{c}$ slopes and stellar ages that is a result of Galactic chemical evolution effects. The youngest stars in the Galactic disk are more overabundant in refractory elements with respect to the Sun and older stars. On the other hand, additional insights by \citet{Ramirez10} and \citet{Melendez12} confirmed the existence of the [X/H]-$T_{c}$ slope as a possible planet formation signature. In agreement with this view, \citet{Gonzalez10} found that stars with planets have more negative [X/H]-$T_{c}$ slopes. Similarly, \citet{Maldonado15} found that the [X/H]-$T_{c}$ slope steepness could also be related to the architecture of the planetary systems: negative slopes have been found for stars with cool giant planets, while stars with hot giant planets or low-mass planets exhibit positive slopes. However, Maldonado and collaborators also cautioned that the Galactic chemical evolution could affect the [X/H]-$T_{c}$ slopes as well. 

Observations of wide binary systems, whose components are two solar analogues, provided further indications that planet formation can play a role in setting the [X/H]-$T_{c}$ slope. The components of a stellar system are most likely formed from the collapse of the same gaseous clump, thus they should share the same composition. High-resolution spectra of the 16~Cyg binary system, composed of a planet-hosting star and a star without detected planets, revealed that the star with the planet is poorer in iron by $\sim$0.04~dex \citep{Ramirez11}. Recently, \citet{TucciMaia14} have shown that this difference in not limited to iron, but involves all the refractory elements. Similar results have been found for the two components of XO-2 \citep{Teske15,Biazzo15,Ramirez15}. We
note, however, that no differences have been found within the error bars for the stars of the binary systems HAT-P-1 \citep{Liu14} and HD~80606 \citep{Saffe15}.

An alternative mechanism that could generate a positive [X/H]-$T_{c}$ slope is represented by a planet engulfment event. The migration of giant planets can induce other rocky objects to move onto unstable orbits and be accreted onto the central star \citep{Sandquist02,Ida08,Zhou08}. After penetrating the stellar atmosphere, the rocky mass would be rapidly dissolved and pollute the ambient matter with refractory elements. Strong hints have been provided by \citet{Schuler11}, who showed that stars with close-in giant planets have positive [X/H]-$T_{c}$ slopes. Recently, \citet{Spina14} found a member of the Gamma Velorum cluster that is significantly enriched in iron relative to the other members of the same cluster, which, as for binary stars, are also assumed to share the same initial compositions. A detailed chemical analysis revealed that the highly refractory elements (those with a $T_{c}$$>$1100~K) are also enhanced by $\sim$0.20~dex \citep{Spina15}, which is most likely the result of the stellar engulfment of $\sim$30~$M_{\oplus}$ of rocky planetary material or planetesimals. 

Similar hints of pollution caused by ingestion of heavy elements come from observations of some white dwarfs that are thought to be accreting material from their planetesimals belts (e.g., \citealt{Gansicke12}).

In addition to observations, numerical estimates \citep{Chambers10,Mack14,Yana15} demonstrated that the stellar ingestion of protoplanetary material or planets could reproduce the positive [X/H]-$T_{c}$ slopes observed in solar twins. Based on the \citet{Siess00} evolutionary models of pre-main-sequence stars, \citet{Spina15} moreover showed that the infall of an extremely large quantity of rocky material equal to that contained in the entire solar system could not produce an iron enhancement greater than 0.01~dex if the accretion episodes occur when the solar-type star is younger than $\sim$10~Myr. However, alternative evolutionary models predict that severe bursty accretion episodes from the circumstellar disk can have the effect of reducing this timescale to a few Myr \citep{Baraffe10}.

Other possible processes have been proposed to explain the different steepnesses of [X/H]-$T_{c}$ slopes observed among the solar
twins. The [X/H]-$T_{c}$ slope could be also affected by a quick photoevaporation of the circumstellar disks that is more likely to occur for stars born in more dense stellar environments or too close to high-mass stars \citep{Onehag14}. Moreover, as suggested by \citet{Gaidos15}, dust-gas segregation in protoplanetary disks can also influence the abundance of refractory elements relative to the volatiles in the stellar atmospheres.

In the context of this lively debate, N15 determined the [X/H]-$T_{c}$ slopes for the solar twins of his sample that have ages spanning $\sim$8 Gyr. From these data he was able to reproduce the correlation between [X/H]-$T_{c}$ slope values and stellar ages observed by \citet{Adibekyan14}, confirming that the Galactic chemical evolution plays a role in setting the [X/H]-$T_{c}$ slopes. On the other hand, he was unable to exclude that other processes, related to the fate of the rocky material orbiting the star during the different phases of its evolution, might also affect the [X/H]-$T_{c}$ slopes. 

The main aim of the present paper is to produce new insights on the origin of the [X/H]-$T_{c}$ slopes and verify whether the [X/Fe]-age correlations identified by N15 can exhaustively explain the diversity of [X/H]-$T_{c}$ slope values observed
so far in solar twins.
To do this, we derived high-precision abundances of 21 elements for a new sample of 14 solar twins to investigate the [X/Fe]-age correlations and to deconvolve the age effect from the [X/H]-$T_{c}$ slopes. In Sect.~\ref{Data} we describe the stellar sample, the spectroscopic observations, and the spectral reduction, while Sect.~\ref{analysis} details the spectral analysis procedure. In Sect.~\ref{discussion} we show our data and discuss our scientific results and, finally, concluding remarks are presented in Sect.~\ref{conclusions}.

%__________________________________________________________________

\section{Spectroscopic observations and data reduction}
\label{Data}
The selection of the 14 solar twins was based on their optical-infrared colours relative to the predicted colours of the Sun by \citet{Ramirez05}. In addition, chromospheric activity, stellar rotation velocities, \textit{Hipparcos} parallaxes, and previous spectroscopic parameters were used. The selected sample contains three planet-hosting stars (HD~13931, HD~95128, and HD~106252), whose planetary parameters are listed in Table~\ref{planets}. One star, HD~45184, is in common with the sample of N15.%Overlaps are not present between our stellar sample and that of N15.

The 14 solar twins were observed using HIRES at the Keck I telescope during the night of January 19-20, 2006. The HIRES spectrograph was employed with the highest resolving power setup (R$\sim$10$^{5}$), which covers $\sim$0.4$-$0.8~$\mu$m on the blue, green, and red chips. Each star has been observed for 1$-$2 minutes. The final signal-to-noise ratios (S/N) are within 160 and 430 at 670~nm, with a median of 230. In addition to the stellar observations, a short exposition of Vesta was performed during the night to acquire a solar spectrum with a S/N of 430. The quality of this data set is lower than that of the N15 spectra, but we benefit from a wider spectral range that allows us to detect more absorption features.

We reduced the spectra with the MAKEE HIRES reduction software v5.4.2\footnote{Different versions of the MAKEE software are available online at \url{http://www.astro.caltech.edu/~tb/makee/}.}. Then we executed the Doppler correction and the continuum normalisation using the IRAF tasks \texttt{dopcor} and \texttt{continuum}, respectively.

\begin{table}
%\vspace{-0.2cm}
\tiny
\begin{center}
\caption{\label{planets} Planet properties}
\begin{tabular}{c|cccc|c} 
\hline\hline 
Planet & Mass & Axis & Period & e & Reference \\
 & [M$_{J}$] & [AU] & [yr] & &  \\ \hline 
HD~13931~b & 1.88 & 5.2 & 11.6 & 0.02 & \citet{Howard10b} \\
HD~95128~b & 2.53 & 2.1 & 3.0 & 0.032 & \citet{Butler96} \\
HD~95128~c & 0.54 & 3.6 & 6.6 & 0.098 & \citet{Fischer02b} \\
HD~95128~d & 1.64 & 11.6 & 38.4 & 0.16 & \citet{Gregory10} \\
HD~106252~b & 7.56 & 2.6 & 4.2 & 0.48 & \citet{Wittenmyer09} \\
\hline\hline 
\end{tabular}
\end{center}
%\vspace{-0.3cm}
\end{table}

%______________________________________________ Spectral analysis
\section{Spectral analysis}
\label{analysis}
Our method is based on the line-by-line differential excitation/ionisation balance analysis relative to the solar spectrum (e.g., \citealt{Melendez12,Melendez14,Bedell14,Ramirez14b}). This approach allowed us to achieve stellar parameters ($T_{eff}$, log~g, and $\xi$) and high-precision abundances of 21 elements (C, O, Na, Al, Si, S, K, Ca, Sc, Ti, V, Cr, Mn, Fe, Co, Ni, Cu, Zn, Sr, Y, and Ba) relative to the Sun itself. The atmospheric parameters and theoretical isochrones were then used to determine the stellar ages and masses. In this section we describe the tools and the procedures adopted for this analysis.

\subsection{Tools}
We employed the master list of atomic transition from \citet{Melendez14}. Whenever possible, we selected the cleanest lines. In the end, we used 99 lines for Fe~I, 18 for Fe~II, and 183 for the other elements.

The equivalent widths (hereafter, EWs) of the absorption features included in the master list were measured using the IRAF \texttt{splot} task. These determinations were performed by hand through Gaussian fits relative to continuum regions selected within windows of $\pm$3$\AA$ centred on the lines of interest. To measure the solar and stellar spectra in exactly the same way, we overplotted the observed solar spectrum on each stellar line and took care to set the same continuum in all the stellar spectra. As shown by \citet{Bedell14}, this approach resolves small differences in abundances ($<$0.01~dex) between different stars if the same instrument and resolution have been adopted for the observations.

For the abundance determinations we adopted the Kurucz (ATLAS9) grid of model atmospheres \citep{Castelli04}. The models were linearly interpolated to obtain a grid of points with the required resolution in the parameter space. Employing the appropriate atmospheric models and the EW measurements, we calculated the local thermodynamic equilibrium (LTE) calculation of the chemical abundances with the 2014 version of MOOG \citep{Sneden73}. 

A Python package called \textit{qoyllur-quipu} or \textit{q2}\footnote{The q2 code, developed by I.R., is available online at \url{https://github.com/astroChasqui/q2}.} was used to determine stellar ages and masses. This code employs the isochrone method, which is a widely used approach to calculate ages and masses of stellar objects. Adopting the Yonsei-Yale isochrones \citep{Yi01,Kim02} and a set of atmospheric parameters (i.e., $T_{eff}$, log~g, and [Fe/H]) with the related uncertainties, \textit{q2} interpolates the set of isochrones to match the input parameters \citep{Ramirez13,Ramirez14b} and calculate the age and mass probability distributions.

\subsection{Stellar parameters}
The achievement of accurate stellar parameters is essential for high-precision abundance determinations. With this aim, we followed the procedure described in \citet{Melendez14}. As a first step, we determined the solar abundances assuming the following solar parameters: $T_{eff}$$=$5777~K, log~g$=$4.44~dex, [M/H]$=$0.00~dex, and $\xi=$1.00~km/s. Then we determined the stellar parameters of the other spectra by computing the iron abundances of the solar twins and, through a line-by-line comparison of the abundances with those of the reference star (the Sun), we computed $\Delta$FeI$_{i}$$=$A$^{*}_{i}$$-$A$^{\odot}_{i}$ for each line, where A$^{*}_{i}$ is the Fe~I abundance relative to the \textit{i}-line. Then we iteratively searched for the three equilibria (excitation, ionisation, and the trend between $\Delta$Fe and log~(EW/$\lambda$)). The iterations were executed with a series of steps starting from a set of initial parameters (i.e., the nominal solar parameters) and arriving at the final set of parameters that simultaneously fulfil the equilibria. We considered the equilibria to be reached when a) the slope in $\Delta$FeI with excitation potential (mainly sensitive to the model $T_{eff}$) is zero within on third  of the slope uncertainty that is due to the scatter in abundances; b) the slope in $\Delta$FeI with log~(EW/$\lambda$) (mainly sensitive to the $\xi$) is zero within one third  of the slope uncertainty; c) the difference $\Delta$FeI-$\Delta$FeII is zero within 1/3$\times$$\sqrt{s.e.^{2}_{FeI}+s.e.^{2}_{FeII}}$ where s.e$.^{2}_{FeI}$ and s.e$.^{2}_{FeII}$ are the standard errors ($\sigma$/$\sqrt{n}$) associated with $\Delta$FeI and $\Delta$FeII, respectively. We also imposed the model metallicity to be equal to that obtained from the iron lines.

\begin{table*}
%\vspace{-0.2cm}
\begin{center}
\begin{threeparttable}
\caption{\label{parameters} Stellar parameters.}
\begin{tabular}{c|cccc|cc|cc} 
\hline\hline 
Star & $T_{eff}$ & log g & [Fe/H] & $\xi$ & Age & Mass & A(Li) & Age$_{Li}$\\
 & [K] & [dex] & [dex] & [km/s] & [Gyr] & [$M_{\sun}$] & [dex] & [Gyr] \\ \hline

HD 9986 & 5827$\pm$7 & 4.44$\pm$0.04 & 0.088$\pm$0.006 & 1.02$\pm$0.01 & 3.3$\pm$1.3 & 1.04$\pm$0.01 & ... & ... \\
HD 13531 & 5653$\pm$12 & 4.53$\pm$0.02 & 0.020$\pm$0.012 & 1.20$\pm$0.03 & 1.8$\pm$1.0 & 0.98$\pm$0.02 & 2.22$\pm$0.02 & 1.6$\pm$0.8 \\
HD 13931 & 5895$\pm$9 & 4.29$\pm$0.02 & 0.067$\pm$0.010 & 1.15$\pm$0.03 & 5.7$\pm$0.4 & 1.07$\pm$0.02 & ... & ... \\
HD 32963 & 5768$\pm$5 & 4.37$\pm$0.03 & 0.088$\pm$0.009 & 0.99$\pm$0.03 & 5.9$\pm$0.8 & 1.03$\pm$0.02 & ... & ... \\
HD 33636 & 5963$\pm$21 & 4.47$\pm$0.04 & $-$0.082$\pm$0.015 & 1.11$\pm$0.03 & 1.7$\pm$1.2 & 1.07$\pm$0.02 & 2.55$\pm$0.02 & 0.9$\pm$0.8 \\
HD 43162 & 5661$\pm$27 & 4.53$\pm$0.03 & 0.057$\pm$0.022 & 1.20$\pm$0.05 & 1.9$\pm$1.3 & 0.99$\pm$0.02 & 2.27$\pm$0.03 & 1.5$\pm$0.8 \\
HD 45184 & 5873$\pm$18 & 4.41$\pm$0.04 & 0.070$\pm$0.016 & 1.03$\pm$0.04 & 3.7$\pm$1.2 & 1.06$\pm$0.02 & ... & ... \\
HD 87359 & 5700$\pm$11 & 4.47$\pm$0.05 & 0.065$\pm$0.009 & 1.05$\pm$0.02 & 4.0$\pm$2.0 & 1.00$\pm$0.02 & ... & ... \\
HD 95128 & 5904$\pm$7 & 4.35$\pm$0.02 & 0.022$\pm$0.006 & 1.14$\pm$0.01 & 5.2$\pm$0.5 & 1.05$\pm$0.01 & ... & ... \\
HD 98618 & 5845$\pm$10 & 4.42$\pm$0.03 & 0.054$\pm$0.009 & 1.05$\pm$0.02 & 3.9$\pm$1.0 & 1.04$\pm$0.01 & ... & ... \\
HD 106252 & 5885$\pm$8 & 4.42$\pm$0.03 & $-$0.069$\pm$0.008 & 1.13$\pm$0.02 & 4.3$\pm$1.1 & 1.02$\pm$0.01 & ... & ... \\
HD 112257* & 5686$\pm$9 & 4.30$\pm$0.02 & $-$0.003$\pm$0.008 & 0.94$\pm$0.02 & 9.6$\pm$0.5 & 0.97$\pm$0.02 & ... & ... \\
HD 140538 & 5704$\pm$13 & 4.48$\pm$0.05 & 0.059$\pm$0.015 & 0.94$\pm$0.03 & 3.6$\pm$1.9 & 1.00$\pm$0.02 & ... & ... \\
HD 143436 & 5825$\pm$14 & 4.43$\pm$0.05 & 0.044$\pm$0.013 & 1.02$\pm$0.03 & 4.2$\pm$1.6 & 1.03$\pm$0.01 & ... & ... \\\hline\hline 
\end{tabular}
\begin{tablenotes}
      \tiny
      \item Note: (*) high-$\alpha$ metal-rich star.     
      \end{tablenotes}
    \end{threeparttable}
\end{center}
%\vspace{-0.3cm}
\end{table*}

We stress that this analysis is strictly differential relative to the Sun, thus any systematic error, such as uncertainties in log~$gf$ and in the continuum setting, should cancel out. Any systematic error in the models that affects the abundances of the reference star (e.g., a not strict fulfilment of the equilibrium conditions using a model with the nominal solar parameters) is likewise identical for the target stars. Thus, because we analyse solar twins, imposing the equilibrium conditions relative to the reference star balances this uncertainty in the stellar abundances \citep{Bedell14}.

We evaluated the errors associated with the stellar parameters following the procedure described in \citet{Epstein10} and \citet{Bensby14}. Since each atmospheric parameter is dependent on the others, this approach takes into account this dependence by propagating the error associated with the fulfilment of the three equilibrium conditions (i.e., minimization of the iron slopes and of $\Delta$FeI$-$$\Delta$FeII) in every single parameter. The quadratic sum of all the error sources associated with each parameter gives its final uncertainty. The typical precisions that we reached are $\sigma$($T_{eff}$)$=$12~K, $\sigma$(log~g)$=$0.03~dex, $\sigma$([Fe/H])$=$0.01~dex, and $\sigma$($\xi$)$=$0.03~km/s.

In the first five columns of Table~\ref{parameters} we list for each star the atmospheric parameters determined by our analysis with their uncertainties. We used these final parameters to calculate the stellar abundances.

\subsection{Chemical abundances}
\label{chemicalabundances}

We employed LTE for the chemical abundance determinations of most of the species using the $\it{abfind}$ driver in the MOOG code. We took into account the hyperfine splitting (HFS) effects that affect the odd-Z elements V, Mn, Co, Cu, Y, and Ba by feeding MOOG with our EWs measurements, but using the $\it{blends}$ driver. This particular driver allows forcing the abundances to match the blended-line EWs when including the hyperfine structures. We assumed the HFS line list adopted by \citet{Melendez14}. All elemental abundances were normalised relative to the solar values on a line-by-line basis. This differential analysis reduces the effect of any systematic error on the abundance determinations of the stellar
parameters. We calculated the NLTE corrections for the oxygen abundances as a function of stellar $T_{eff}$, log~g, and [Fe/H] using an IDL script based on the NLTE corrections computed by \citet{Ramirez07}. We note that the absolute NLTE effects are of the order of $\sim$0.15~dex, but the differential NLTE effects are no more than 0.04~dex.

We calculated the error budget associated with the abundances [X/H] following a procedure similar to that described in \citet{Melendez12}, who took two types of uncertainty sources into account: the observational error that is due to the line-to-line scatter in the EW measurements, and the errors in the atmospheric parameters. If more than one line is available for a given element, we assumed the observational error to be the standard error. When, as for potassium, only one line is detected, the error was estimated by repeating the EW measurement five times with different assumptions on the continuum setting and taking the standard deviation ($\sigma$) of the resulting abundances. In addition, the uncertainties related to the errors in the atmospheric parameters were estimated by summing quadratically the abundance changes produced by the alteration of every single parameter by its own error. Then, the observational error and that related to the stellar parameters were quadratically summed, which resulted in the final uncertainty that affects the abundance. For the [X/Fe] ratios, the uncertainties related to the errors in the parameters were estimated from the changes in the elemental abundance produced by alterations in the parameters relative to the changes in iron produced by the same alterations. A similar approach was used for the [C/O] ratios.

Titanium and chromium were also observed in two ionisation states. For these two we adopted as final abundances the classical weighted averages among the abundances derived from the neutral and first ionised states, as follows:
$$\widehat{A}=\frac{\sum A_{i}/\sigma^{2}_{i}}{\sum 1/\sigma^{2}_{i}},$$
where A$_{i}$ is the abundance in one of the two states and $\sigma$$_{i}$ is the relative uncertainty.
We consistently estimated the error using the formula
$$\sigma (\widehat{A})=\frac{1}{\sqrt{\sum 1/\sigma^{2}_{i}}}.$$
The same procedure was used to calculate the carbon abundance that was also observed in the CH molecular state. %Also scandium has been detected in both the states, but we used the Sc~II only since it is less uncertain than Sc~I.

The [X/H] determinations that resulted from the chemical analysis described above with the related errors are listed in Table~\ref{XH_ratios}, while the [X/Fe] ratios are listed in Table~\ref{XFe_ratios}.

%______________________________________________ Stellar ages
\subsection{Stellar ages}
We fed the \textit{q2} code with the set of atmospheric parameters and relative uncertainties listed in Table~\ref{parameters} to derive ages and masses for the sample of solar twins. For each star q2 computed two probability distributions for the two parameters we aim to determine. From these distributions we took the confidence interval at the 65$\%$ level as the limits in age and masses. The age assigned to each solar twin was assumed to be the centre value of that interval. Consequently, we took the half-width of the interval as the age error bar. Analogously, we determined the stellar masses and their uncertainties from the mass probability distributions. These values are listed in Cols. 6 and 7 of Table~\ref{parameters}.

\begin{figure*}
\centering
\includegraphics[width=18cm]{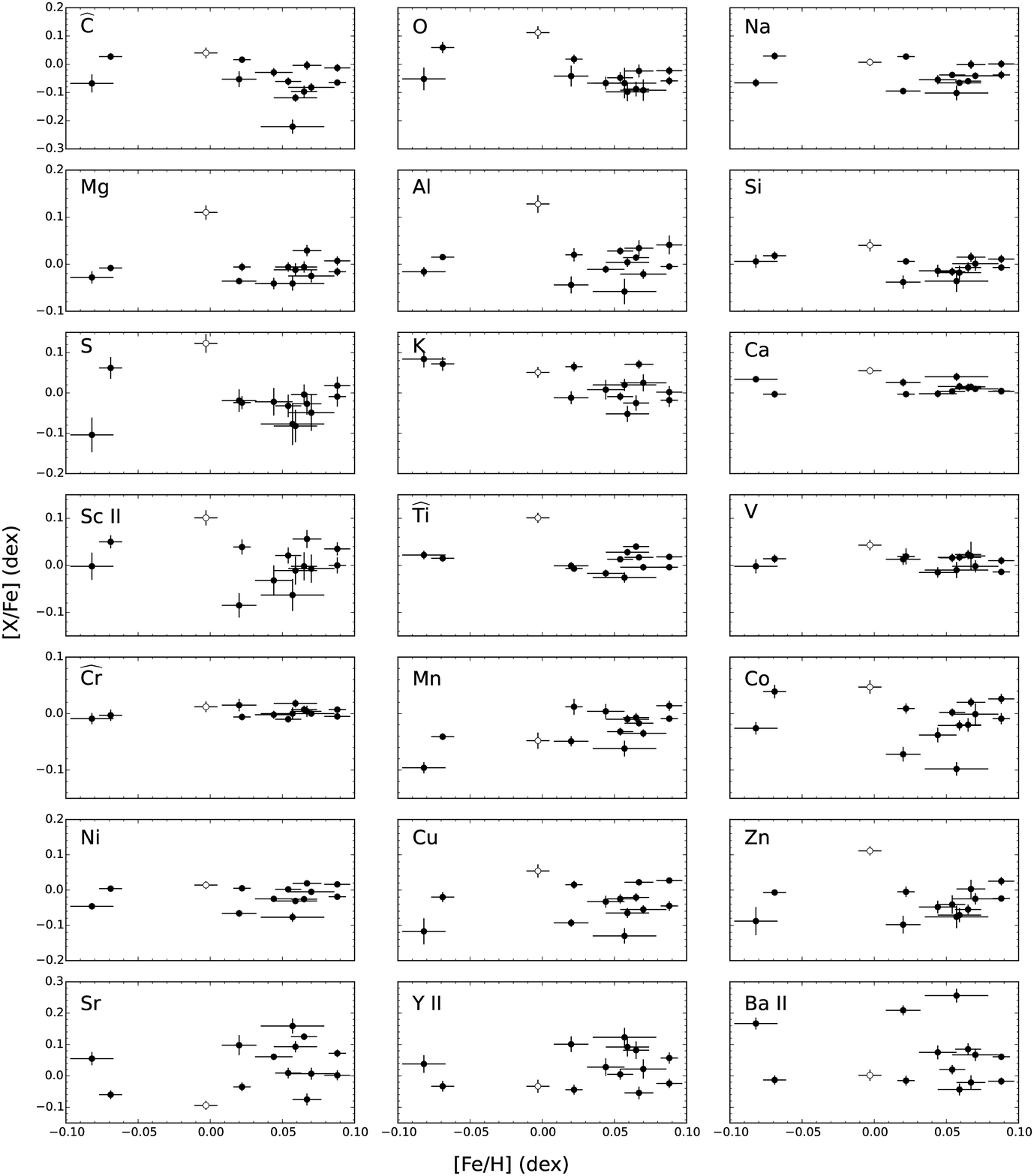} \caption{[X/Fe] values as a function of stellar metallicities [Fe/H]. The abundances and corresponding errors are those listed in Table~\ref{XFe_ratios}. The empty circle is the h$\alpha$mr star HD~112257, while the filled circles are the 13 normal thin-disk stars of our sample.}
\label{XFe_FeH}
\end{figure*}

In addition to the ages calculated by q2, we also performed more precise age determinations from the lithium abundances A(Li) of the three youngest stars in our sample (i.e., HD 13531, HD 33636, and HD 43162). In fact, even though the lithium doublet features at 6708$\AA$ are detectable in many spectra of our sample, it is particularly strong in these three
spectra and appropriate for a reliable determination of the elemental abundance. Furthermore, considering the youth of the three stars, the A(Li) is an excellent age tracer for these three solar twins, since the element is being burned faster than later during the main-sequence. As a first step, we therefore measured the rotational velocities v$\sin$i on five iron lines (i.e., 6027.050, 6093.644, 6151.618, 6165.360, and 6705.102 $\AA$) and one Ni line at 6767.772 $\AA \text{ for each star} $. These lines were chosen for their proximity to the Li doublet, because their features are not contaminated by blends, and because of their intermediate strength (40$\lesssim$EW$\lesssim$80~m$\AA$). Given the Fe and Ni abundances listed in Table~\ref{XH_ratios}, we used the $\it{synth}$ drive in the MOOG code to compute the synthetic spectra of the six lines, and by varying the v$sin$i parameter, we aimed to reproduce \textit{} their observed profiles
by eye. We adopted the average of the six v$\sin$i determinations as a fixed parameter to determine the A(Li) that could reproduce the observed Li doublet feature. We also assumed the laboratory parameters of the Li doublet and the adjacent lines, both atomic and molecular,  employed by \citet{Melendez12} for their A(Li) determinations. The A(Li) uncertainties were estimated by i) varying the A(Li) abundance to reproduce the observed feature; ii) taking different assumptions on the spectral continuum setting; iii) taking into account the errors in the atmospheric parameters. The quadratic sum of the three error sources yields uncertainties that are smaller than 0.03~dex. The A(Li) determinations are reported in Col. of Table~\ref{parameters}. Given these A(Li) values, we calculated the stellar ages assuming the models from \citet{Andrassy15}. \citet{Carlos15} have shown that stars with A(Li)$\sim$2.2~dex have ages falling within $\pm$0.8~Gyr, thus we conservatively adopted this value as error bar. Our age determinations obtained from A(Li) are listed in the last column of Table~\ref{parameters}.

In Sect.~\ref{analysis} we determined the stellar parameters, ages, masses, and abundances for the whole sample of solar twins. 
Our sample has one star in common (i.e., HD~45184) with the sample of N15, which allows us to validate our results. For this particular star, the agreement between our measurements and those of N15 is within one sigma with very high confidence given that only two ([Fe/H] and [Ni/Fe]) out of 19 measurements have differences that are marginally over the one-sigma errors.

%______________________________________________ Discussion
\section{Discussion}
\label{discussion}
In this section, we discuss the relations between the high-precision abundances studied in solar-twin stars and their metallicities and ages. These correlations have previously been investigated by N15 for some of the elements he studied (i.e., C, O, Na, Mg, Al, Si, S, Ca, Ti, Cr, Ni, Zn, and Y). We compare our results with those found by N15 and expand the current knowledge for K, Sc, V, Mn Co, Cu, Sr, and Ba. We also examine the chemical patterns of our targets, focusing on the relation between the chemical abundances and the condensation temperature (hereafter, $T_{c}$). Finally, we explore how this relation developed with time as a result of the chemical evolution of the Galactic thin-disk
stars.

\subsection{Relations between elemental abundances and metallicity}
\label{abu_metal}
Iron is an excellent proxy of the stellar metallicity because
of the metals present in the stellar atmospheres, it is one of the greatest contributors in terms of mass. In addition to this, optical spectra are rich in iron lines, which is useful for a reliable abundance determination. We therefore used this element to explore any possible dependence between the chemical abundances we derived for the sample of solar twins and their metallicities. In Fig. \ref{XFe_FeH} we show the [X/Fe] abundances as a function of [Fe/H]. Within the range of metallicities covered by our sample ([Fe/H]$\sim$$\pm$0.10~dex), most of the elements have a [X/Fe] scatter that is larger than their error bars. The only exceptions are Ca, Cr, and V, which show no dependence on [Fe/H].

\begin{figure}
\centering
\includegraphics[width=8.5cm]{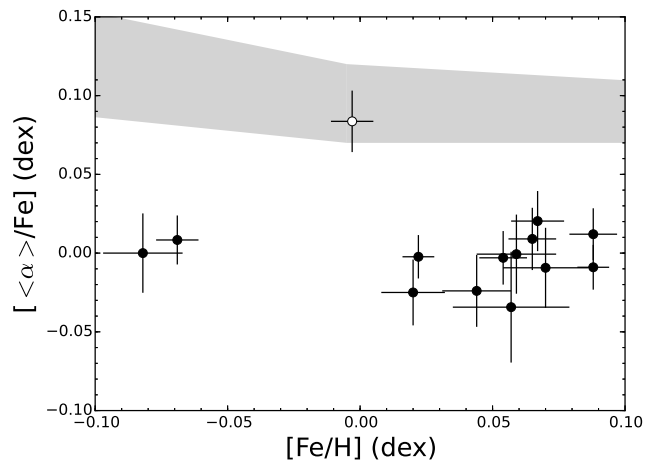} \caption{[$<$$\alpha$$>$/Fe] versus [Fe/H] for the whole sample of solar twins. For $<$$\alpha$$>$ we mean the average of the Mg, Si and Ti abundances. The symbols are the same as in Fig.~\ref{XFe_FeH}, where the empty circle corresponds to the HD~112257 star. The grey band highlights the typical range of [$<$$\alpha$$>$/Fe] values found for the h$\alpha$mr stars by \citet{Adibekyan11}.}
\label{alpha_FeH}
\end{figure}

We also noted that the star HD~112257 (empty circle) is systematically enriched in most of the $\alpha$-elements (i.e., O, Mg, Si, S, and Ti) with respect to the other stars. Of the $\alpha$-elements, only Ca is not significantly overabundant in this star. In addition, HD~112257 is also the richest in C, Al, Sc, Cu, and Zn. Given the parallax of this object (23.39~mas; \citealt{vanLeeuwen07}), the proper motions ($\mu_{\alpha}$$=-$11.16, $\mu_{\delta}$$=-$71.49~mas/yr; \citealt{vanLeeuwen07}), and its radial velocity ($-$39.323~km/s; \citealt{Soubiran13}), HD~112257 has a probability of 97$\%$ of belonging to the thin-disk population according to \citet{Reddy06}. However, the chemical pattern described above is typical of the high-$\alpha$ metal-rich (h$\alpha$mr) population first identified by \citet{Adibekyan11}. This class of stars is kinematically indistinguishable from the thin-disk population \citep{Adibekyan12b,Bensby14}, but they are characterised by greater ages (i.e., $\sim$8-10~Gyr; \citealt{Haywood13}) and higher [$<$$\alpha$$>$/Fe] values ($\sim$0.1~dex) with respect the normal thin-disk stars that have ages up to $\sim$8~Gyr and [$<$$\alpha$$>$/Fe]$\sim$0.0~dex at solar metallicities. In Fig. \ref{alpha_FeH} we plot the [$<$$\alpha$$>$/Fe] ratios (where $<$$\alpha$$>$ is the average of Mg, Si, and Ti, according to the plots in \citealt{Adibekyan12b}) as a function of iron content. The plot shows that the solar-twins have a [$<$$\alpha$$>$/Fe] average of 0.00~dex with a standard dispersion $\sigma$=0.03~dex, while HD~112257 is a 3$\sigma$ outlier with a [$<$$\alpha$$>$/Fe]=0.08$\pm$0.02~dex, which is perfectly compatible with the typical [$<$$\alpha$$>$/Fe] values of the h$\alpha$mr stars found by Adibekyan et al. (grey band). Moreover, \citet{Adibekyan12b} showed that h$\alpha$mr stars are on average also over-abundant in Al, Sc, Co, and Ca with respect to the thin-disk population, which is consistent with what we found for HD~112257. We therefore conclude that this object belongs to the h$\alpha$mr population and treat it separately from the sample of normal thin-disk stars.

Interestingly, the scatter is still present for most of the elements shown in Fig. \ref{XFe_FeH} regardless of the h$\alpha$mr star HD~112257. Obviously, this can be attributed to the fact that many other factors in addition to the simple stellar metallicity could affect the abundance of each individual element. 

\begin{figure}
\centering
\includegraphics[width=8.5cm]{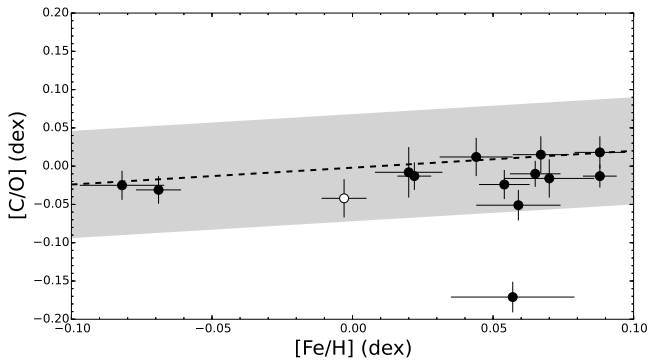} \caption{[C/O] ratios as a function of [Fe/H] for the whole sample of solar twins. The symbols are the same as in Fig.~\ref{XFe_FeH}, where the empty circle corresponds to the HD~112257 star. The linear relation between the [C/O] ratios and [Fe/H] fitted by N15 is plotted as a dashed line, while the grey band represents the one-sigma dispersion around the fit using the data set of \citet{Takeda05}. The star with the anomalously low C abundance is HD~43162.}
\label{C_O}
\end{figure}

In Fig.~ \ref{C_O} we plot the [C/O] ratios as a function of [Fe/H] for all the solar twins we analysed. The graph shows that our sample contains one star (i.e., HD~43162) with a strikingly low [C/O] ratio of $-$0.17$\pm$0.05~dex. Recently \citet{Nissen13}, using high-precision C and O abundances from \citet{Takeda05}, found that the [C/O] ratios of thin-disk solar-type stars follow the linear relation plotted in Fig.~ \ref{C_O} as a dashed line with a one-sigma confidence interval of $\pm$0.07~dex (grey band). Similarly, he found a 0.058 scatter from the \citet{Ramirez09} [C/O] determinations. Using the \citet{Takeda05} data set, we calculate a fraction of 7$\%$ of stars with [C/O]$\leq-$0.17~dex within the metallicity range spanned by our sample. It is generally accepted that the oxygen mainly comes from massive stars that
exploded in type II SNe, while the origin of carbon is more uncertain. Like oxygen, carbon is also produced by explosions of massive stars, but it is still debated if its principal contribution comes from winds of low- or intermediate-mass stars, such as \textit{AGB} stars (e.g., \citealt{Akerman04,Gavilan05,Mattsson10}). Interestingly, from Fig.~\ref{C_O} we note that HD~43162 has the lowest [C/Fe] ratio, while its [O/Fe] value is consistent with those of many other solar twins. %Considering that, apart from the [C/Fe] ratio, HD~43162 does not show any other chemical peculiarity, not even for the $\alpha$-elements, it is likely that this star formed in an environment poorly polluted of carbon by low- and intermediate-mass stars.
The origin of this uncommon carbon abundance is still unclear and demands further dedicate observations.

\subsection{Chemical evolution of the Galactic thin-disk stars}
\begin{figure*}
\centering
\includegraphics[width=18cm]{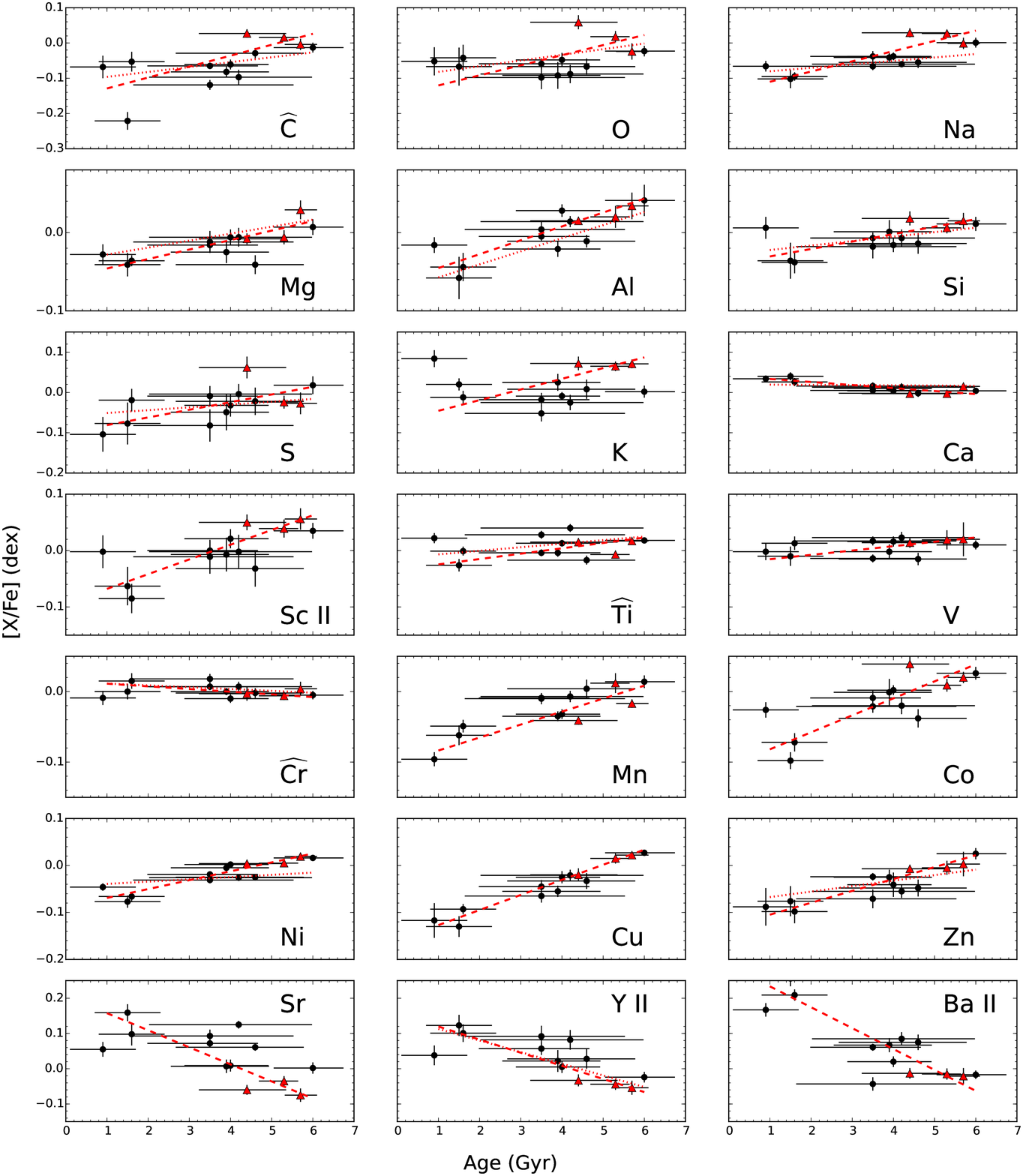} \caption{[X/Fe] ratios as a function of stellar ages for the 13 normal thin-disk stars of our sample. Red filled triangles highlight the three planet-hosting stars. The red dashed lines correspond to the linear fits, whose coefficients are listed in Table~\ref{age_fit}. We also overplot the linear correlations found by \citet{Nissen15} with red dotted lines.}
\label{age_trend}
\end{figure*}

As shown by N15, tight correlations between [X/Fe] and age exist for most of the species analysed in his study. These relations can be partially explained by the rising fraction between type Ia and type II supernovae with time. In Fig.~\ref{age_trend} we show these relations for the species we detected in our sample of 13 normal thin-disk solar-twin stars (we excluded HD~112257 from the plots since it belongs to a separate population). For the plot we adopted the age estimates listed in Table~\ref{parameters} and, in particular for the stars HD~13531, HD~33636, and HD~43162, we assumed the ages derived from A(Li). Using $\textit{MPFIT}$ \citep{Markwardt09}, an IDL routine able to perform least-squares fittings that take into account the uncertainties in both abscissa and ordinate, we performed a linear fit of these correlations (i.e., [X/Fe]$=$Age$\times$b+c). We did not include HD~43162 data in the fit of the carbon abundance vs age, since that star has an anomalous [C/Fe] ratio, as we showed in Sect.~\ref{abu_metal}. The resulting b and c parameters and their relative errors are listed in Table \ref{age_fit} together with the standard deviations of the residuals ($\sigma_{[X/Fe]}$) and the $\chi^{2}_{red}$. This latter is the typical chi-square value weighted for the measurement error vectors in the two coordinates.% The last column of Table \ref{age_fit} reports the average of the uncertainties associated to the abundances of the X element calculated among the stars considered for the fit ($<$$\sigma_{X}$$>$). 

The linear functions that resulted from our fits are plotted in Fig. \ref{age_trend} as dashed red lines, while the dotted lines represent the relations found by N15, when available. 
We observe that our stellar sample has a general behaviour that resembles the one described by N15: the stars are aligned in relations between [X/Fe] and ages with amplitudes up to 0.30~dex, like for Ba. In some cases, such as for Ca and Cr, these relations are extremely tight ($\sigma_{[X/Fe]}$$<$0.01). In other cases they are much more scattered, as for O, S, K, Sr, and Ba that have $\sigma_{[X/Fe]}$$\geq$0.05. The C and Ti distribution are also scattered relative to the abundance uncertainties, since their $\chi^{2}_{red}$ values are 1.40 and 3.51, respectively. The high dispersion level observed for K is expected, since its abundances are only based on the saturated line at 7698.97$\AA$, so that very small variations in the EW measurements cause great changes in the elemental abundance. On the other hand, the abundances of the other elements, such as Ti, are much more reliable, and the observed dispersion around the linear fit might therefore be a real effect that is due to different chemical histories. However, all these more scattered elements also show a clear correlation with age, and the higher dispersion is mainly due to the presence of one or several outliers. Our sample shows
no systematic outlier from the main trends. The three planet-hosting stars (red filled triangles) have the same behaviour as all the other stars of our sample without any strong and systematic chemical anomaly.

Interestingly, most of the elements show positive correlations between [X/Fe] ratios and ages. The only exceptions are the three neutron-capture elements (i.e., Sr, Ba, and Y), whose [X/Fe] ratios decrease with age. This result agrees with the one found by N15 for Y and strengthens his suggestion that these anti-correlations might be due to the delayed production of s-process elements by low-mass AGB stars.

%\begin{figure}
%\centering
%\includegraphics[width=8.5cm]{/Users/lspina/Copy/final_spectra/plots/figures/Fe_age.eps} \caption{Bla bla bla.}
%\label{Fe_age}
%\end{figure}

\begin{table}
%\vspace{-0.2cm}
\tiny
\begin{center}
\caption{\label{age_fit} Results of the linear fit: [X/Fe] $=$ Age $\times$ b + a}
\begin{tabular}{c|cc|cc} 
\hline\hline 
Element & a & b & $\sigma_{[X/Fe]}$ & $\chi^{2}_{red}$ \\
 & [dex] & [10$^{-2}$dex/Gyr] & [dex] & \\ \hline 
C & $-$0.160$\pm$0.031 &  3.10$\pm$0.66 & 0.043 & 1.40  \\
O & $-$0.149$\pm$0.032 &  2.86$\pm$0.70 & 0.049 & 1.70  \\
Na & $-$0.139$\pm$0.022 &  2.90$\pm$0.48 & 0.027 & 0.96  \\
Mg & $-$0.058$\pm$0.011 &  1.21$\pm$0.28 & 0.013 & 0.67 \\
Al & $-$0.063$\pm$0.017 &  1.77$\pm$0.41 & 0.016 & 0.65  \\
Si & $-$0.040$\pm$0.012 &  0.95$\pm$0.27 & 0.014 & 0.93  \\
S & $-$0.100$\pm$0.030 &  1.90$\pm$0.65 & 0.034 & 1.22  \\
K & $-$0.072$\pm$0.024 &  2.65$\pm$0.55 & 0.056 & 3.69 \\
Ca & 0.041$\pm$0.008 & $-$0.75$\pm$0.19 & 0.008 & 0.82 \\
Sc & $-$0.094$\pm$0.026 &  2.61$\pm$0.59 & 0.028 & 0.72 \\
Ti & $-$0.034$\pm$0.010 &  0.95$\pm$0.22 & 0.020 & 3.51 \\
V & $-$0.023$\pm$0.011 &  0.77$\pm$0.27 & 0.013 & 1.04 \\
Cr & 0.015$\pm$0.007 & $-$0.38$\pm$0.16 & 0.009 & 1.07 \\
Mn & $-$0.102$\pm$0.014 &  1.84$\pm$0.32 & 0.017 & 1.15 \\
Co & $-$0.106$\pm$0.018 &  2.41$\pm$0.40 & 0.025 & 1.24 \\
Ni & $-$0.088$\pm$0.014 &  1.89$\pm$0.30 & 0.013 & 0.48 \\
Cu & $-$0.159$\pm$0.025 &  3.21$\pm$0.52 & 0.012 & 0.12 \\
Zn & $-$0.130$\pm$0.026 &  2.53$\pm$0.58 & 0.018 & 0.23  \\
Sr & 0.206$\pm$0.037 & $-$4.84$\pm$0.83 & 0.057 & 1.28 \\
Y2 & 0.157$\pm$0.030 & $-$3.72$\pm$0.69 & 0.040 & 0.78 \\
Ba2 & 0.293$\pm$0.039 & $-$5.93$\pm$0.89 & 0.049 & 0.62 \\
\hline\hline 
\end{tabular}
\end{center}
%\vspace{-0.3cm}
\end{table}

We also note that Mg, Al, Si, Ca, Ti, Cr, and Y have slope values that are consistent or very similar to those provided by N15. On the other hand, the slopes found for C, O, Na, S, Ni, and Zn are significantly greater than the slopes reported by N15. Interestingly, this difference is systematic and might partially be attributed to the different age ranges spanned by the two samples: the N15 sample covers ages from $\sim$1 to $\sim$8 Gyr, while our sample reaches up to $\sim$6 Gyr. Some of the elemental distributions plotted in Fig. 8 of N15 show that most of the oldest stars ($\gtrsim$7~Gyr) of that sample tend to lie below the fitted correlation, which drives the slopes toward lower values. This behaviour is not general to all the elements, but it is particularly evident for Na and Ni. As a test, we fitted the N15 Na and Ni abundances excluding the stars with ages $\geq$7~Gyr. As a result, we obtained slopes that are consistent with our values. This suggests that the correlations between [X/Fe] and age are more complex than a linear fit if stars older than $\sim$7 Gyr are included, as in the N15 sample. 
Using the N15 data set and the \citet{Ramirez14} samples of metal-rich solar analogues and late F-type dwarfs, \citet{Ramirez15b} have drawn similar conclusions for the evolution of Na and Ni. This non-linear behaviour could also explain the relatively high $\chi^{2}_{red}$ values that N15 derived for Na and Ni (i.e., $\chi^{2}_{red}>$10), regardless of the very tight correlation between [Na/Fe] and [Ni/Fe] that he identified for solar-twin stars. 
%\textbf{[NOTE: Nissen wrote that the amount of Na and Ni produced by Type II SNe depends on the neutron excess, which itself depends on metallicity and the $\alpha$/Fe. Is it possible that this dependence caused the lower Na and Ni abundances at old ages observed in Nissen data?]} 
For S and Zn we also obtained smaller $\chi^{2}_{red}$ and $\sigma_{[X/Fe]}$ than the N15 values. This points to a generalised non-linear behaviour for stars older than 7~Gyr. The case of C and O is
different: the high scatter of their distributions observed both from our and the N15 data could account for the different slopes.

\subsection{ [X/H]-$T_{c}$ slopes}
\label{The[X/H]-$T_{c}$ slopes}
As mentioned in the introduction, several studies on solar twins have shown that their chemical patterns are characterised by their own [X/H]-$T_{c}$ slope. In this section, we intend to study the [X/H]-$T_{c}$ slopes of our sample of solar twins to investigate their possible origins and their relation with the chemical evolution of the Galactic thin-disk stars. With this aim, we performed a linear least-squares fitting of the [X/H] vs $T_{c}$ relation for each star, taking into account the uncertainties of our abundance estimates and assuming the $T_{c}$ values from \citet{Lodders03}. The neutron-capture elements (i.e., Sr, Y and Ba) were not included in the fitting, since their dependence on the $T_{c}$ is in general more complex than a linear fit, as we discuss below. The resulting coefficients are listed in Table~\ref{Tcond} together with their uncertainties, the average of the residuals $\sigma_{[X/H]}$ , and the $\chi^{2}_{red}$ of the fits. We note that our [X/H]-$T_{c}$ slope for HD~45184 is consistent with that measured by N15: the difference between the two measurements is 0.5$\pm$1.4~10$^{-5}$~dex~K$^{-1}$. The stellar chemical patterns are shown in Figs. \ref{Tcond_trends_1} and \ref{Tcond_trends_2}, where the [X/H]-$T_{c}$ slopes are also represented.

\begin{table}
\tiny
%\vspace{-0.2cm}
\begin{center}
\begin{threeparttable}
\caption{\label{Tcond} [X/H]-T$_{c}$ slope coefficients. [X/H]$=$T$_{c}$$\times$b + a}
\begin{tabular}{c|cc|cc} 
\hline\hline 

Star & a & b & $\sigma_{[X/H]}$ & $\chi^{2}_{red}$ \\
 & [dex] & [10$^{-5}$dex K$^{-1}$] & [dex] & \\ \hline 
HD 9986 & 0.028 $\pm$ 0.007 & 3.75 $\pm$ 0.52 & 0.012 & 1.61 \\
HD 13531 & $-$0.082 $\pm$ 0.015 & 4.81 $\pm$ 1.15 & 0.040 & 6.75 \\
HD 13931 & 0.051 $\pm$ 0.012 & 2.50 $\pm$ 0.90 & 0.020 & 2.03 \\
HD 32963 & 0.078 $\pm$ 0.009 & 1.79 $\pm$ 0.69 & 0.012 & 1.22 \\
HD 33636 & $-$0.175 $\pm$ 0.015 & 5.81 $\pm$ 1.13 & 0.044 & 4.45 \\
HD 43162 & $-$0.101 $\pm$ 0.019 & 8.63 $\pm$ 1.48 & 0.049 & 3.98 \\
HD 45184 & $-$0.010 $\pm$ 0.014 & 5.05 $\pm$ 1.03 & 0.020 & 1.50 \\
HD 87359 & $-$0.047 $\pm$ 0.010 & 7.92 $\pm$ 0.78 & 0.018 & 2.68 \\
HD 95128 & 0.047 $\pm$ 0.008 & $-$1.38 $\pm$ 0.61 & 0.020 & 3.19 \\
HD 98618 & $-$0.026 $\pm$ 0.009 & 5.64 $\pm$ 0.71 & 0.010 & 1.07 \\
HD 106252 & $-$0.055 $\pm$ 0.007 & $-$0.16 $\pm$ 0.51 & 0.029 & 5.22 \\
HD 112257* & 0.065 $\pm$ 0.010 & $-$1.39 $\pm$ 0.76 & 0.047 & 16.89 \\
HD 140538 & $-$0.082 $\pm$ 0.013 & 9.35 $\pm$ 0.97 & 0.018 & 1.46 \\
HD 143436 & $-$0.022 $\pm$ 0.011 & 3.44 $\pm$ 0.78 & 0.018 & 1.05 \\

\hline\hline 
\end{tabular}
\begin{tablenotes}
      \tiny
      \item Note: (*) high-$\alpha$ metal-rich star.     
      \end{tablenotes}
    \end{threeparttable}
\end{center}
%\vspace{-0.3cm}
\end{table}

\begin{figure*}
\centering
\includegraphics[width=18cm]{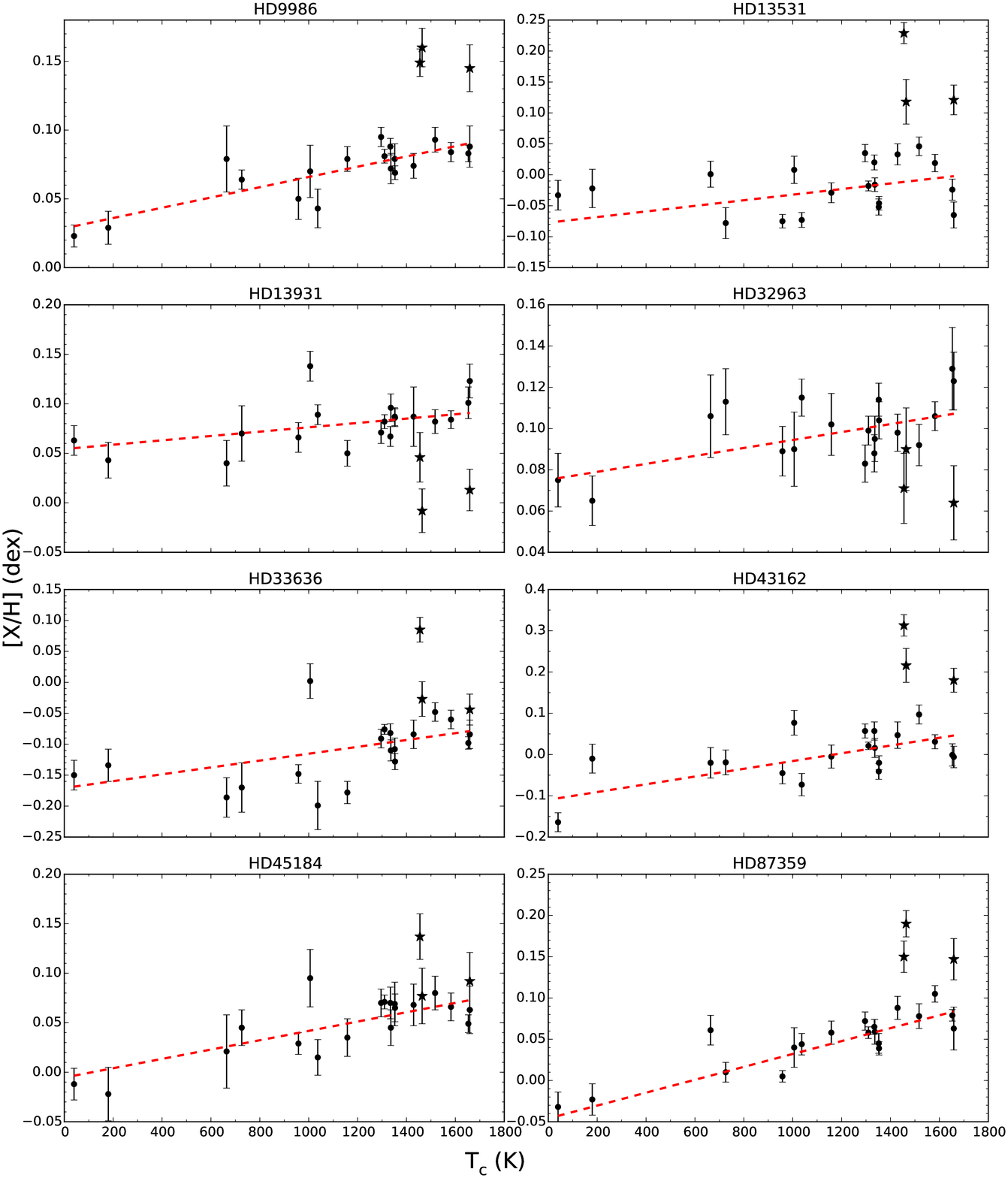} \caption{Chemical patterns of each single solar-twin star. Abundances [X/H] are plotted as a function of condensation temperature $T_{c}$. Light elements (Z$\leq$30) are represented by filled circles, while stars are used for the heavy elements (Z$>$30). The [X/H]-$T_{c}$ slopes, whose coefficients are listed in Table~\ref{Tcond}, are shown as red dashed lines.}
\label{Tcond_trends_1}
\end{figure*}

\begin{figure*}
\centering
\includegraphics[width=18cm]{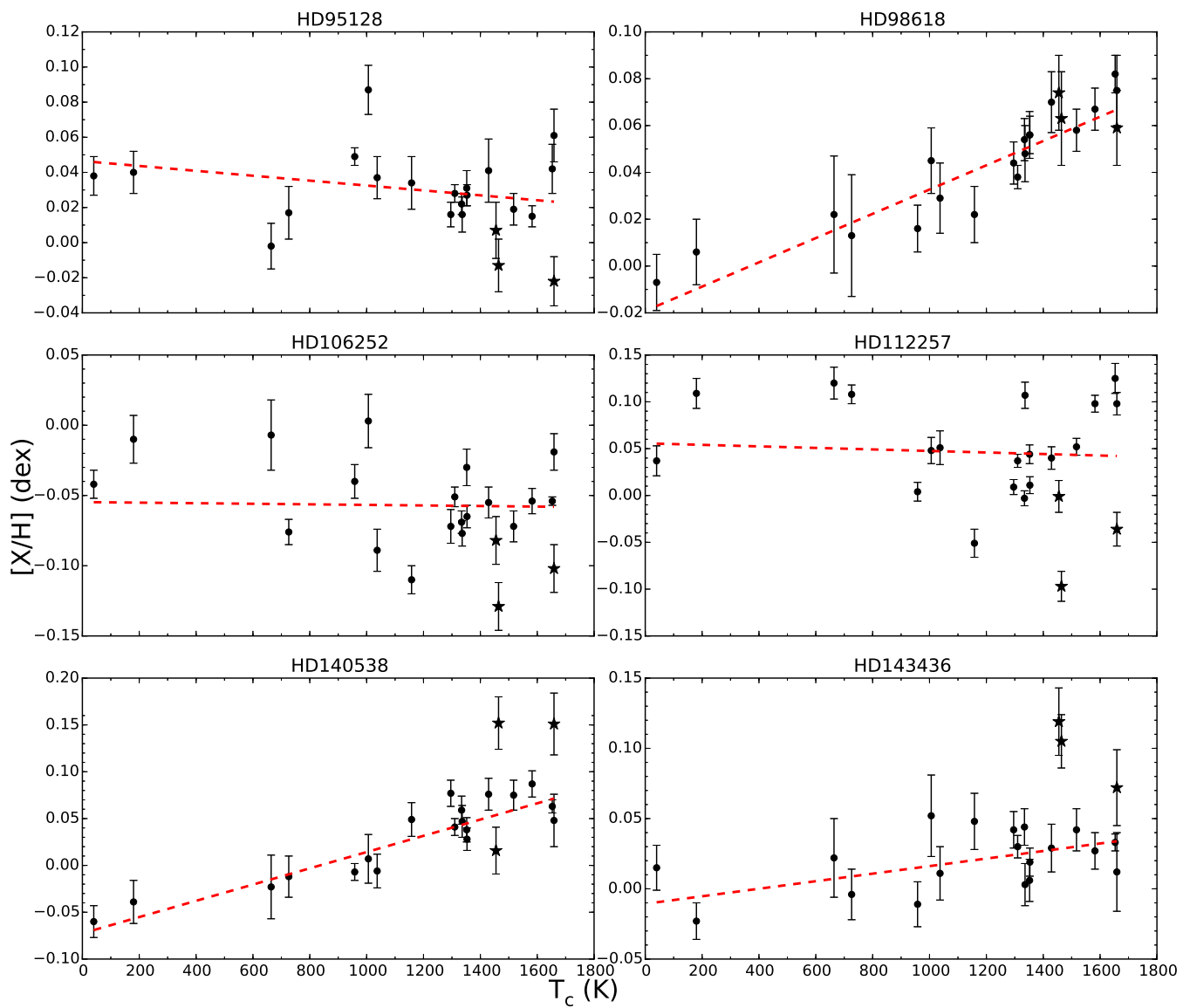} \caption{Continuation of Fig.~\ref{Tcond_trends_1}}
\label{Tcond_trends_2}
\end{figure*}

As Table~\ref{Tcond} shows, the linear fits of our data yield a $\chi^{2}_{red}$$>$1.0 for all the stars. In Figs. \ref{Tcond_trends_1} and \ref{Tcond_trends_2} the residuals from the fit for many elements are clearly greater than their uncertainties. The average dispersion around the fitted [X/H]-$T_{c}$ slopes calculated for the normal thin-disk stars is $<$$\sigma_{[X/H]}$$>=$0.024~dex, while the average of the goodness-of-fit indicator is $<$$\chi^{2}_{red}$$>=$2.79. N15 also pointed out that he was unable to find a perfect correlation between the elemental abundances and their $T_{c}$ in his sample
(he also derived $\chi^{2}_{red}$$>$1.0 for all the stars). However, we also note that in both our sample and in that of N15 some solar-twin stars have very steep [X/H]-$T_{c}$ slopes, with values several times higher than their uncertainties (e.g., HD~87359 or HD~140538). These considerations indicate that, indeed, each chemical pattern of thin-disk stars is roughly outlined by its own [X/H]-$T_{c}$ slope value. Nevertheless, there are also some second-order chemical departures that mark each stellar content and produce the observed dispersion. An exemplifying case of these deviations is represented by the neutron-capture elements. For most of the stars, the abundances greatly differ from the main trend that is delineated by the volatiles and refractory elements. This different behaviour was already observed by \citet{Melendez14} in the solar twin 18~Sco, which appears enriched in neutron-capture elements with respect to with its [X/H]-$T_{c}$ relation relative to the Sun. They explained this complex abundance pattern of the heavy elements as a sign of pollution from s-process and r-process ejecta with respect to that experienced by the Sun, which is considerably older than 18~Sco. Consistently, \citet{Ramirez11} found that the two members of 16~Cyg, which are older than the Sun by about 2.5~Gyr, have neutron-capture elements that are
lower than their [X/H]-$T_{c}$ relations. N15 also noted that all solar-twin stars he analysed exhibit an additional enhancement in neutron-capture elements with respect to their [X/H]-$T_{c}$ slopes. However, the heavy elements are not always enhanced since Figs. \ref{Tcond_trends_1} and \ref{Tcond_trends_2} clearly show that in five solar twins in our sample Sr, Y, and Ba are significantly lower than the [X/H]-$T_{c}$ relations.

From Table~\ref{Tcond} we also observe that about 80$\%$ of our stars have positive [X/H]-$T_{c}$ slopes. Only HD~106262 has a slope consistent with zero, while two stars (i.e., HD~95128 and HD~112257) exhibit a slightly negative slope. This result confirms that compared to the majority of the solar-twin stars, the Sun has a peculiar chemical pattern characterised by higher abundances of the volatile elements and/or lower abundances of the refractories. Analysing a sample of 11 solar-twins, \citet{Melendez09} concluded similarly: 90$\%$ of the stars show a strictly positive [X/H]-$T_{c}$ slope. This also agrees with the results of N15,  who obtained an occurrence of 80$\%$ for the whole sample and 95$\%$ when the h$\alpha$mr stars are excluded. However, it should be noted that the chemical pattern of the h$\alpha$mr star HD~112257 is more complex than those of normal thin-disk solar twins, since its linear fit of the [X/H]-$T_{c}$ relation gave the highest $\chi^{2}_{red}$ value (i.e., $\chi^{2}_{red}$$=$16.89).  Figure \ref{Tcond_trends_2} shows that its chemical pattern its more characterised by a systematic over-abundance of $\alpha$-elements (e.g., O, Mg, S, and Ti) and of several other elements (e.g., Al, Sc, and Zn), and has no real correlation with $T_{c}$.

\begin{figure}
\centering
\includegraphics[width=8.5cm]{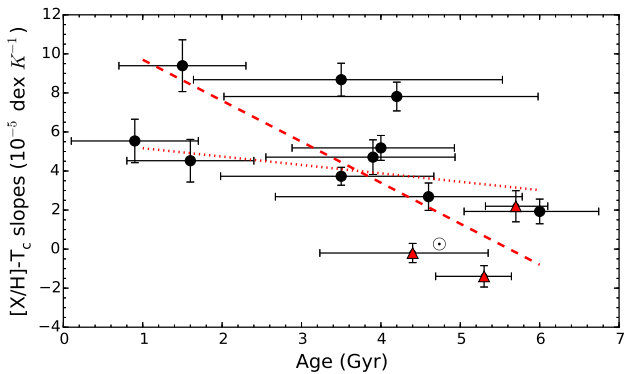} \caption{[X/H]-$T_{c}$ slopes as a function of stellar ages. The red filled triangles are the planet-hosting stars. The red dashed line is the result of the linear fit of the data set shown in the plot, while the red dotted line represents the [X/H]-$T_{c}$ slopes vs age relation found by \citet{Nissen15}.}
\label{Tcond_age}
\end{figure}

The relation between the [X/H]-$T_{c}$ slopes and stellar ages has been unveiled by a recent study of \citet{Ramirez14} and then confirmed by \citet{Adibekyan14}. N15 was also able to reproduce the same behaviour, finding a similar age dependence of the [X/H]-$T_{c}$ slope to that described by Adibekyan et al.: the N15 sample has been fitted by a $-$0.43$\pm$0.15~10$^{-5}$~dex~K$^{-1}$~Gyr$^{-1}$ relation. In Fig.~\ref{Tcond_age} we show the [X/H]-$T_{c}$ slopes listed in Table~\ref{Tcond} as a function of stellar ages. We fitted this data set through a linear least-squares fitting and
obtained the relation
$$[X/H]-T_{c}~slope=11.8(\pm 1.7)-2.1(\pm 0.4) \times Age[Gyr], $$
where the X/H]-$T_{c}$ slope is in 10$^{-5}$ dex K$^{-1}$ units. This fitting gave a $\chi^{2}_{red}$$=$2.1. The relation slope that we found (shown in Fig.~\ref{Tcond_age} as a dashed line) is much lower than that of N15 (dotted line in Fig.~\ref{Tcond_age}), and the two values are not consistent within the errors. However, we recall that the age range covered by our solar twins is limited to 6~Gyr and is about 2 Gyr shorter than that of the N15 sample. In addition, the [X/H]-$T_{c}$ slopes that we obtained from our sample span very different values, from $-$1.38 to +9.35~10$^{-5}$~dex~K$^{-1}$, while the N15 values are restricted to the +0.4~-~+6.7~10$^{-5}$~dex~K$^{-1}$ range. Interestingly, two of the three stars with planets (i.e., HD~95128 and HD~106252) and the Sun have the lowest [X/H]-$T_{c}$ slope values. However, this could be a mere effect of the relation between [X/H]-$T_{c}$ slopes and ages, since these stars lie on the right side of the diagram in Fig.~\ref{Tcond_age}, corresponding to the oldest stars of the sample.

\subsection{Subtracting the chemical evolution effect}

In the previous paragraphs we have shown that the elemental abundances of each solar-twin star tend to be aligned in clear relations with $T_{c}$. Consequently, the [X/H]-$T_{c}$ slope value is a distinguishing feature that characterises the chemical pattern of each normal thin-disk star. In addition to this, the dispersion of the elemental abundances around the [X/H]-$T_{c}$ slopes cannot be completely ascribed to the uncertainties affecting the abundance values. A wide range of [X/H]-$T_{c}$ slope values has also been observed at all the ages in the plot of Fig.~\ref{Tcond_age}. It is not still completely clear to which degree these chemical departures from the well-established relations discussed in Sect.~\ref{The[X/H]-$T_{c}$ slopes} might be attributed to the chemical evolution of the Galactic thin-disk stars, or whether they are exclusively the signature of other processes, such as pollution from stellar ejecta and SNe explosions, which randomly affect the observed chemical patterns. Moreover, it is still not proven if the relation between [X/H]-$T_{c}$ slope values and the stellar ages found by several authors (e.g., \citealt{Adibekyan14,Ramirez14,Nissen15}) and also shown in Fig.~\ref{Tcond_age} is purely a consequence of the relations between [X/Fe] and ages plotted in Fig.~\ref{age_trend}. To address all these open questions, we subtracted the effect of the chemical evolution from the stellar chemical patterns by defining for each X element a ``residual abundance'' as follows
$$\left[ \frac{X}{Fe} \right]_{res}=\left[ \frac{X}{Fe} \right]_{obs}-\left[ \frac{X}{Fe} \right]_{fit,}$$
where the [X/Fe]$_{obs}$ is the observed abundance and [X/Fe]$_{fit}$ the abundance expected from the stellar age assuming the fits listed in Table~\ref{age_fit}. Therefore, we calculated the [X/Fe]$_{res}$ for the abundances of all the 14 solar-twins. We also calculated the [X/Fe]$_{res}$  for the Sun, assuming a solar age of 4.6~Gyr and [X/Fe]$_{obs}$$=$0~dex.
 
\begin{figure}
\centering
\includegraphics[width=8.5cm]{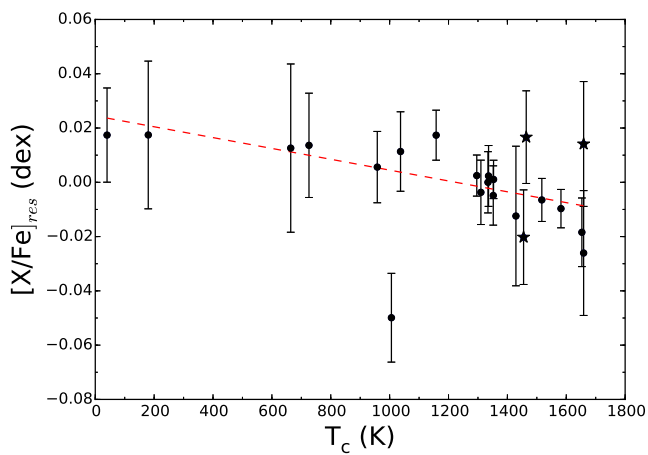} \caption{\textit{Herathin} chemical pattern of the Sun. The [X/Fe]$_{res}$ calculated for the Sun are plotted as a function of  $T_{c}$. The red dashed line represents the \textit{herathin} [X/H]-$T_{c}$ slope. Symbols and colours are the same as in Fig. \ref{Tcond_trends_1}.}
\label{Tcond_sun}
\end{figure}

 Hereafter we call this set of abundances corrected for the chemical evolution effect \textit{herathin}, an adjective that we invented for this specific scope and that originates from merging the words \textit{Hera} and \textit{thin}$\footnote{According to ancient Greek mythology, the Milky Way originated from Hera, the divine wife of Zeus. Legend has it  that Zeus begot a son called Heracles from a mortal woman. Aiming to provide this
son with godlike powers, Zeus let Hera nurse Heracles while she was asleep. When Hera woke up and realized she was feeding an unknown baby, she pushed him away. This caused some of the milk to splash into the sky and become the Milky Way. Since the matter that formed the Milky Way originated from Hera, the adjective \textit{herathin} characterises something related to the original and pristine matter of the Galactic thin-disk stars.}$. In Fig.~\ref{Tcond_sun} we plot the chemical pattern of the Sun, after subtracting the chemical evolution effect: the \textit{herathin} ``solar'' composition appears slightly enhanced in volatiles and poorer in refractory elements with respect to the actual solar composition. The error bars in this plot represent the typical uncertainties affecting each element from our data: they are the average of all the error estimates of a given element. The \textit{herathin} [X/H]-$T_{c}$ slope of the Sun is $-$2.59$\pm$0.96~10$^{-5}$~dex~K$^{-1}$. All the elements are nearly consistent within their errors with the fit (shown in Fig.~\ref{Tcond_sun} as a dashed red line) except from K, but this element has the most uncertain [X/Fe] vs age fit from Table \ref{age_fit}. 

\begin{table}
\tiny
%\vspace{-0.2cm}
\begin{center}
\begin{threeparttable}
\caption{\label{age_fit_res} \textit{Herathin} [X/H]-T$_{c}$ slope coefficients. [X/H]$_{res.}=$T$_{c}$$\times$b + a}
\begin{tabular}{c|cc|cc} 
\hline\hline 

Star & a & b & $\sigma_{[X/H]}$ & $\chi^{2}_{red}$ \\
 & [dex] & [10$^{-5}$dex K$^{-1}$] & [dex] & \\ \hline 
HD 9986 & 0.090 $\pm$ 0.007 & $-$0.13 $\pm$ 0.52 & 0.015 & 1.97 \\
HD 13531 & 0.067 $\pm$ 0.015 & $-$3.56 $\pm$ 1.15 & 0.016 & 1.07 \\
HD 13931 & 0.031 $\pm$ 0.012 & 2.56 $\pm$ 0.90 & 0.010 & 0.66 \\
HD 32963 & 0.051 $\pm$ 0.009 & 2.16 $\pm$ 0.69 & 0.021 & 1.97 \\
HD 33636 & $-$0.036 $\pm$ 0.015 & $-$1.38 $\pm$ 1.13 & 0.036 & 3.08 \\
HD 43162 & 0.025 $\pm$ 0.019 & 1.80 $\pm$ 1.48 & 0.033 & 1.38 \\
HD 45184 & 0.043 $\pm$ 0.014 & 1.51 $\pm$ 1.03 & 0.016 & 1.15 \\
HD 87359 & $-$0.003 $\pm$ 0.010 & 4.74 $\pm$ 0.77 & 0.022 & 2.63 \\
HD 95128 & 0.041 $\pm$ 0.008 & $-$1.85 $\pm$ 0.61 & 0.011 & 1.30 \\
HD 98618 & 0.017 $\pm$ 0.009 & 2.69 $\pm$ 0.71 & 0.012 & 1.48 \\
HD 106252 & $-$0.020 $\pm$ 0.007 & $-$3.06 $\pm$ 0.51 & 0.021 & 2.74 \\
HD 112257* & 0.064 $\pm$ 0.010 & $-$1.47 $\pm$ 0.76 & 0.047 & 17.73 \\
HD 140538 & $-$0.014 $\pm$ 0.013 & 5.16 $\pm$ 0.97 & 0.019 & 1.10 \\
HD 143436 & 0.006 $\pm$ 0.011 & 0.99 $\pm$ 0.78 & 0.019 & 1.45 \\
Sun & 0.025 $\pm$ 0.012 & $-$2.14 $\pm$ 0.93 & 0.014 & 0.57 \\

\hline\hline 
\end{tabular}
\begin{tablenotes}
      \tiny
      \item Note: (*) high-$\alpha$ metal-rich star.     
      \end{tablenotes}
    \end{threeparttable}
\end{center}
%\vspace{-0.3cm}
\end{table}

In Table~\ref{age_fit_res} we list the zero points and slope coefficients obtained from the linear least-squares fitting of the \textit{herathin} chemical patterns, together with the dispersion around the fit and the $\chi^{2}_{red}$. Interestingly, a consequence of the correction for the chemical evolution is the general decrease of $\sigma_{[X/H]}$ and $\chi^{2}_{red}$. The averages of the dispersion in abundances ($<$$\sigma_{[X/H]}$$>$) of all the normal thin-disk stars passed from 0.024~dex (in Table~\ref{age_fit}) to 0.019 (in Table~\ref{age_fit_res}). Similarly, the average of the $\chi^{2}_{red}$ decreased to 1.69 from the initial value of 2.79 in Table~\ref{age_fit}. This suggests that the scatter around the [X/H] vs $T_{c}$ correlations observed in Figs.~\ref{Tcond_trends_1} and \ref{Tcond_trends_2} is partially due to the chemical evolution of the Galactic disk.

\begin{figure}
\centering
\includegraphics[width=8.5cm]{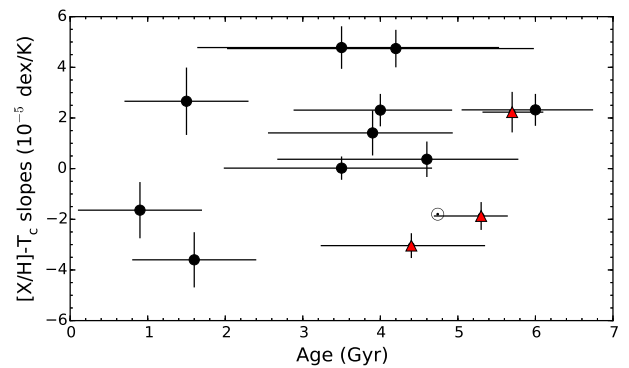} \caption{\textit{Herathin} [X/H]-$T_{c}$ slopes as a function of stellar ages. Symbols and colours as in Fig.~\ref{Tcond_age}.}
\label{Tcond_age_norm}
\end{figure}

Another important result is that, after correcting for the age effect, only eight stars have a strictly positive [X/H]-$T_{c}$ slope, one has a slope consistent with zero, and four stars and the Sun show a negative [X/H] vs $T_{c}$ correlation. We plot in Fig.~\ref{Tcond_age_norm} these [X/H]-$T_{c}$ slope values for the normal thin-disk stars and for the Sun as a function of  age. Interestingly, subtracting the chemical evolution effect in the stellar chemical patterns completely removes the [X/H]-$T_{c}$ slope vs age anticorrelation. On the other hand, the wide range of [X/H]-$T_{c}$ slope values (from $-$4~10$^{-5}$ to $+$4~10$^{-5}$~dex~K$^{-1}$) at all the ages is still present in the diagram. This strong diversity of the \textit{herathin} [X/H]-$T_{c}$ slopes indicates that the chemical evolution of the Galactic disk can only partially affect the [X/H]-$T_{c}$ slope steepness exhibited by each single star. In addition to a development with time, other mechanisms must play a role in differentiating the [X/H]-$T_{c}$ slopes. %The different capability of circumstellar disks to form planets could reasonably account for the variety of these abundance patterns that we observed in Fig.~\ref{Tcond_age_norm} \citep{Melendez09,Ramirez09,Chambers10}, rather than the pollution from AGB stars or supernovae explosions \citep{Melendez12}, which, contrarily, are the main responsible of the Galactic chemical evolution. Support to this hypothesis is given by the fact that the 60$\%$ of the stars in our sample with a [X/H]-$T_{c}$ slope~$<$~0~dex K$^{-1}$ show evidences of harbouring planets (i.e., HD106252, HD 95128 and the Sun), while among the nine solar-twins with [X/H]-$T_{c}$ slope~$\geq$~0~dex K$^{-1}$ there is only one star with a known planetary system (i.e., HD13931). Obviously, a larger statistic is needed to confirm if planet-hosting stars are segregated in the diagram shown in Fig.~\ref{Tcond_age_norm}.

%
%______________________________________________________________

\section{Summary and conclusions}
\label{conclusions}
We derived high-precision chemical abundances of 22 elements from the HIRES spectra of 14 solar-twin stars. This data set allowed us to confirm that [X/Fe] ratios of all the species correlate with age (see Fig.~\ref{age_trend}). These dependences reflect the chemical evolution of the Galactic disk and have been also studied by N15 for C, O, Na, Mg, Al Si, S, Ca, Ti, Cr, Ni, Zn, and Y. With the present paper we extended our knowledge also to K, Sc, V, Mn, Co, Cu, Sr, and Ba. The linear fitting of these correlations resulted in slopes that are consistent with those of N15 for seven species, but our slope values of C, O, Na, S, Ni, and Zn are systematically greater than those of N15. In particular, this difference for Na and Ni may be due to the wider age range spanned by the N15 sample and to a non-linear behaviour of abundances from stars older than $\sim$7~Gyr.

The abundance determinations were also useful to derive the [X/H]-$T_{c}$ slopes for all the solar twins shown in Fig.~\ref{Tcond_trends_1}. We confirmed that each chemical pattern of the normal thin-disk stars in our sample is characterised by a distinctive relation between abundances and $T_{c}$. Nevertheless, we note that these dependences have a larger scatter than expected: the fits resulted in  $<$$\sigma_{[X/H]}$$>$$=$0.024~dex and a $<$$\chi^{2}_{red}$$>$$=$2.79. We also saw that about 80$\%$ of the stars in our sample show a positive slope, and we confirmed that the [X/H]-$T_{c}$ slope values decrease with time (see Fig.~\ref{Tcond_age}). On the other hand, the h$\alpha$mr star HD~112257 has a completely different chemical pattern that cannot be outlined by a [X/H]-$T_{c}$ slope.
We employed the [X/Fe]-age relations to subtract the evolution effect from the chemical patterns of the solar twins. As shown in Fig.~\ref{Tcond_age_norm}, this correction reduced the average scatter around the unevolved (\textit{herathin}) [X/H]-$T_{c}$ slopes by 0.005~dex, meaning that the chemical evolution introduced a scatter of 0.01~dex and decreased the $<$$\chi^{2}_{red}$$>$ by 1.10. It also removed any dependence between the \textit{herathin} [X/H]-$T_{c}$ slopes and age. 

Most interestingly, we observed that the \textit{herathin} [X/H]-$T_{c}$ slopes are highly dispersed at all the ages within a wide range of values (i.e., $\pm$4~10$^{-5}$~dex~K$^{-1}$). This diversity indicates that in addition to the chemical evolution of the Galactic thin-disk stars, other processes may contribute to setting the [X/H]-$T_{c}$ slopes. As we recalled in the introduction, a variety of events could account for different modifications of the stellar [X/H]-$T_{c}$ slopes. The formation of planetary systems would lead to a preferential accretion of volatile elements, while refractories would be locked in the rocky bodies orbiting the star. Similarly, a dust-gas segregation in protoplanetary disks could induce a difference in the mass of volatiles (relative to the refractories) that is accreted onto the star. Planetary engulfment episodes would also enrich the stellar photosphere of refractory elements. On the other hand, other processes may favour a preservation of the original chemical pattern without altering the abundance of refractory elements relative to the volatiles in significant quantities (e.g., the failure of planet formation or a quick photo-evaporation of the circumstellar disk).

In the past two decades, we passed from the discovery of the first extrasolar planet around 51~Peg \citep{Mayor95} to the discovery of distant Jupiters like the Jupiter in our solar system \citep{Bedell15}. Increasing efforts have been made to detect new exo-planets and study their nature, orbits, and the characteristics of their hosting stars. Even though we are not completely aware whether all the solar-type stars are capable of forming planets or not, the current picture shows that the planetary systems can be very different in the number of planets, types of planets, and system architectures. In addition, it has been shown by several authors that most of the exo-planets discovered so far did not form in their current configuration. Observations of Jupiters with extremely small orbits or eccentricities up to those of the cometary objects that populate our solar system clearly indicate that several planetary systems undergo processes of orbital reconfiguration and planet migration (e.g., \citealt{Morbidelli11}, and references therein) during the first few hundred Myr of their existence. The origin of this great difference between the solar system and known exo-planetary systems or among the exo-planetary systems themselves lies in these highly dynamic stages of planetary systems. During this chaotic development of the system's architecture, part of the rocky material orbiting the star (e.g., cores of gaseous planets, rocky planets, and planetesimals) may be induced to move onto unstable orbits and fall onto the central star. On the other hand, other planetary systems can survive without any dramatic modification of their structure \citep{Izidoro15}. According to this view, the wide range of [X/H]-$T_{c}$ slope values spanned at all the ages by our sample of solar-twin stars could reflect this variety of fates that the matter in circumstellar disks and the planetary systems can experience.

\begin{acknowledgements}
L.S. and J.M acknowledge the support from FAPESP (2014/15706-9 and 2012/24392-2).
\end{acknowledgements}

% WARNING
%-------------------------------------------------------------------
% Please note that we have included the references to the file aa.dem in
% order to compile it, but we ask you to:
%
% - use BibTeX with the regular commands:
%   \bibliographystyle{aa} % style aa.bst
%   \bibliography{Yourfile} % your references Yourfile.bib
%
% - join the .bib files when you upload your source files
%-------------------------------------------------------------------

\bibliographystyle{aa}
\bibliography{/Users/lspina/Copy/papers/bibliography.bib}

\newpage
\newpage
\clearpage

\begin{sidewaystable*}
\begin{threeparttable}
\caption{ [X/H] abundance ratios}\label{XH_ratios}
        \centering 
\tiny
\begin{tabular}{l|cccccccccccc} 
\hline\hline             
HD & C & CH & $\widehat{\textrm{C}}$ & O & Na & Mg & Al & Si & S & K & Ca & Sc II  \\ \hline
   9986 &  0.016$\pm$0.011 &  0.033$\pm$0.013 &  0.023$\pm$0.008 &  0.029$\pm$0.012 &  0.050$\pm$0.015 &  0.072$\pm$0.011 &  0.083$\pm$0.006 &  0.081$\pm$0.005 &  0.079$\pm$0.024 &  0.070$\pm$0.019 &  0.093$\pm$0.009 &  0.088$\pm$0.015 \\
   13531 & -0.030$\pm$0.029 & -0.040$\pm$0.046 & -0.033$\pm$0.024 & -0.022$\pm$0.031 & -0.075$\pm$0.011 & -0.016$\pm$0.011 & -0.024$\pm$0.017 & -0.018$\pm$0.008 &  0.001$\pm$0.021 &  0.008$\pm$0.022 &  0.046$\pm$0.015 & -0.065$\pm$0.021 \\
   13931 &  0.058$\pm$0.022 &  0.068$\pm$0.020 &  0.063$\pm$0.015 &  0.043$\pm$0.018 &  0.066$\pm$0.015 &  0.096$\pm$0.014 &  0.101$\pm$0.016 &  0.082$\pm$0.007 &  0.040$\pm$0.023 &  0.138$\pm$0.015 &  0.082$\pm$0.012 &  0.123$\pm$0.017 \\
   32963 &  0.083$\pm$0.018 &  0.067$\pm$0.019 &  0.075$\pm$0.013 &  0.065$\pm$0.012 &  0.089$\pm$0.012 &  0.095$\pm$0.011 &  0.129$\pm$0.020 &  0.099$\pm$0.007 &  0.106$\pm$0.020 &  0.090$\pm$0.018 &  0.092$\pm$0.010 &  0.123$\pm$0.014 \\
   33636 & -0.159$\pm$0.025 & -0.088$\pm$0.065 & -0.150$\pm$0.024 & -0.134$\pm$0.026 & -0.148$\pm$0.015 & -0.110$\pm$0.017 & -0.098$\pm$0.010 & -0.076$\pm$0.008 & -0.186$\pm$0.032 &  0.002$\pm$0.028 & -0.048$\pm$0.015 & -0.084$\pm$0.023 \\
   43162 & -0.181$\pm$0.030 & -0.142$\pm$0.034 & -0.164$\pm$0.023 & -0.010$\pm$0.035 & -0.045$\pm$0.026 &  0.016$\pm$0.023 & -0.001$\pm$0.027 &  0.021$\pm$0.009 & -0.020$\pm$0.037 &  0.077$\pm$0.030 &  0.097$\pm$0.023 & -0.006$\pm$0.026 \\
   45184 & -0.038$\pm$0.024 &  0.009$\pm$0.021 & -0.012$\pm$0.016 & -0.022$\pm$0.027 &  0.029$\pm$0.011 &  0.045$\pm$0.018 &  0.049$\pm$0.009 &  0.071$\pm$0.007 &  0.021$\pm$0.037 &  0.095$\pm$0.029 &  0.080$\pm$0.017 &  0.063$\pm$0.024 \\
   87359 & -0.033$\pm$0.023 & -0.029$\pm$0.032 & -0.032$\pm$0.018 & -0.023$\pm$0.019 &  0.005$\pm$0.007 &  0.059$\pm$0.015 &  0.079$\pm$0.007 &  0.058$\pm$0.007 &  0.061$\pm$0.018 &  0.040$\pm$0.024 &  0.078$\pm$0.015 &  0.063$\pm$0.026 \\
   95128 &  0.027$\pm$0.017 &  0.045$\pm$0.014 &  0.038$\pm$0.011 &  0.040$\pm$0.012 &  0.049$\pm$0.005 &  0.016$\pm$0.010 &  0.042$\pm$0.014 &  0.028$\pm$0.005 & -0.002$\pm$0.013 &  0.087$\pm$0.014 &  0.019$\pm$0.009 &  0.061$\pm$0.015 \\
   98618 & -0.018$\pm$0.016 &  0.011$\pm$0.020 & -0.007$\pm$0.012 &  0.006$\pm$0.014 &  0.016$\pm$0.010 &  0.048$\pm$0.012 &  0.082$\pm$0.008 &  0.038$\pm$0.005 &  0.022$\pm$0.025 &  0.045$\pm$0.014 &  0.058$\pm$0.009 &  0.075$\pm$0.015 \\
   106252 & -0.041$\pm$0.011 & -0.046$\pm$0.019 & -0.042$\pm$0.010 & -0.010$\pm$0.017 & -0.040$\pm$0.012 & -0.077$\pm$0.009 & -0.054$\pm$0.003 & -0.051$\pm$0.007 & -0.007$\pm$0.025 &  0.003$\pm$0.019 & -0.072$\pm$0.011 & -0.019$\pm$0.013 \\
   112257* &  0.067$\pm$0.023 &  0.011$\pm$0.022 &  0.037$\pm$0.016 &  0.109$\pm$0.016 &  0.004$\pm$0.010 &  0.107$\pm$0.014 &  0.125$\pm$0.016 &  0.037$\pm$0.007 &  0.120$\pm$0.017 &  0.048$\pm$0.014 &  0.052$\pm$0.009 &  0.098$\pm$0.012 \\ 
   140538 & -0.090$\pm$0.029 & -0.045$\pm$0.021 & -0.060$\pm$0.017 & -0.039$\pm$0.023 & -0.007$\pm$0.009 &  0.047$\pm$0.017 &  0.063$\pm$0.007 &  0.041$\pm$0.009 & -0.023$\pm$0.034 &  0.007$\pm$0.026 &  0.075$\pm$0.016 &  0.048$\pm$0.028 \\
   143436 & -0.011$\pm$0.026 &  0.029$\pm$0.019 &  0.015$\pm$0.016 & -0.023$\pm$0.013 & -0.011$\pm$0.016 &  0.003$\pm$0.015 &  0.033$\pm$0.006 &  0.030$\pm$0.008 &  0.022$\pm$0.028 &  0.052$\pm$0.029 &  0.042$\pm$0.015 &  0.012$\pm$0.028 \\ \hline \hline
HD & Ti I &  Ti II & $\widehat{\textrm{Ti}}$ & V & Cr I & Cr II & $\widehat{\textrm{Cr}}$ & Mn & Co & Ni  & Cu & Zn \\ \hline
   9986 &  0.084$\pm$0.008 &  0.084$\pm$0.016 &  0.084$\pm$0.007 &  0.074$\pm$0.009 &  0.099$\pm$0.008 &  0.081$\pm$0.016 &  0.095$\pm$0.007 &  0.079$\pm$0.009 &  0.079$\pm$0.011 &  0.069$\pm$0.005 &  0.043$\pm$0.014 &  0.064$\pm$0.007 \\
   13531 &  0.025$\pm$0.015 & -0.019$\pm$0.039 &  0.019$\pm$0.014 &  0.033$\pm$0.017 &  0.046$\pm$0.017 &  0.019$\pm$0.022 &  0.035$\pm$0.014 & -0.029$\pm$0.016 & -0.052$\pm$0.013 & -0.046$\pm$0.011 & -0.073$\pm$0.012 & -0.078$\pm$0.025 \\
   13931 &  0.079$\pm$0.011 &  0.095$\pm$0.016 &  0.084$\pm$0.009 &  0.087$\pm$0.030 &  0.078$\pm$0.014 &  0.060$\pm$0.018 &  0.071$\pm$0.011 &  0.050$\pm$0.013 &  0.087$\pm$0.009 &  0.086$\pm$0.009 &  0.089$\pm$0.010 &  0.070$\pm$0.028 \\
   32963 &  0.108$\pm$0.009 &  0.100$\pm$0.015 &  0.106$\pm$0.007 &  0.098$\pm$0.009 &  0.085$\pm$0.011 &  0.079$\pm$0.015 &  0.083$\pm$0.009 &  0.102$\pm$0.015 &  0.114$\pm$0.008 &  0.104$\pm$0.009 &  0.115$\pm$0.009 &  0.113$\pm$0.016 \\
   33636 & -0.053$\pm$0.019 & -0.071$\pm$0.024 & -0.060$\pm$0.015 & -0.084$\pm$0.023 & -0.090$\pm$0.018 & -0.094$\pm$0.029 & -0.091$\pm$0.015 & -0.178$\pm$0.018 & -0.108$\pm$0.018 & -0.128$\pm$0.013 & -0.199$\pm$0.039 & -0.170$\pm$0.040 \\
   43162 &  0.058$\pm$0.029 &  0.017$\pm$0.021 &  0.031$\pm$0.017 &  0.047$\pm$0.032 &  0.073$\pm$0.026 &  0.045$\pm$0.022 &  0.057$\pm$0.017 & -0.005$\pm$0.028 & -0.041$\pm$0.019 & -0.020$\pm$0.017 & -0.073$\pm$0.027 & -0.019$\pm$0.030 \\
   45184 &  0.067$\pm$0.017 &  0.065$\pm$0.025 &  0.066$\pm$0.014 &  0.068$\pm$0.021 &  0.070$\pm$0.017 &  0.070$\pm$0.025 &  0.070$\pm$0.014 &  0.035$\pm$0.019 &  0.069$\pm$0.022 &  0.065$\pm$0.014 &  0.015$\pm$0.018 &  0.045$\pm$0.018 \\
   87359 &  0.111$\pm$0.012 &  0.079$\pm$0.024 &  0.105$\pm$0.010 &  0.088$\pm$0.014 &  0.075$\pm$0.012 &  0.059$\pm$0.024 &  0.072$\pm$0.011 &  0.058$\pm$0.014 &  0.045$\pm$0.012 &  0.039$\pm$0.008 &  0.044$\pm$0.013 &  0.010$\pm$0.012 \\
   95128 &  0.007$\pm$0.007 &  0.036$\pm$0.012 &  0.015$\pm$0.006 &  0.041$\pm$0.018 &  0.016$\pm$0.008 &  0.017$\pm$0.014 &  0.016$\pm$0.007 &  0.034$\pm$0.015 &  0.031$\pm$0.010 &  0.027$\pm$0.006 &  0.037$\pm$0.012 &  0.017$\pm$0.015 \\
   98618 &  0.063$\pm$0.011 &  0.076$\pm$0.015 &  0.067$\pm$0.009 &  0.070$\pm$0.013 &  0.039$\pm$0.011 &  0.056$\pm$0.017 &  0.044$\pm$0.009 &  0.022$\pm$0.012 &  0.056$\pm$0.010 &  0.056$\pm$0.008 &  0.029$\pm$0.015 &  0.013$\pm$0.026 \\
   106252 & -0.062$\pm$0.011 & -0.037$\pm$0.015 & -0.054$\pm$0.009 & -0.055$\pm$0.011 & -0.072$\pm$0.013 & -0.071$\pm$0.029 & -0.072$\pm$0.012 & -0.110$\pm$0.010 & -0.030$\pm$0.013 & -0.065$\pm$0.008 & -0.089$\pm$0.015 & -0.076$\pm$0.009 \\
   112257* &  0.098$\pm$0.012 &  0.097$\pm$0.015 &  0.098$\pm$0.009 &  0.040$\pm$0.012 &  0.019$\pm$0.010 & -0.020$\pm$0.016 &  0.009$\pm$0.008 & -0.051$\pm$0.015 &  0.044$\pm$0.010 &  0.011$\pm$0.009 &  0.051$\pm$0.018 &  0.108$\pm$0.010 \\ 
   140538 &  0.095$\pm$0.017 &  0.065$\pm$0.027 &  0.087$\pm$0.014 &  0.076$\pm$0.017 &  0.080$\pm$0.016 &  0.066$\pm$0.029 &  0.077$\pm$0.014 &  0.049$\pm$0.018 &  0.038$\pm$0.013 &  0.028$\pm$0.012 & -0.006$\pm$0.018 & -0.012$\pm$0.022 \\
   143436 &  0.028$\pm$0.015 &  0.024$\pm$0.027 &  0.027$\pm$0.013 &  0.029$\pm$0.017 &  0.045$\pm$0.015 &  0.031$\pm$0.027 &  0.042$\pm$0.013 &  0.048$\pm$0.020 &  0.006$\pm$0.015 &  0.019$\pm$0.010 &  0.011$\pm$0.019 & -0.004$\pm$0.018 \\ \hline \hline

\cline{1-4} 

\cline{1-4}
HD & Sr & Y II & Ba \\ \cline{1-4}
      9986 &  0.160$\pm$0.014 &  0.145$\pm$0.017 &  0.149$\pm$0.010 & \multicolumn{9}{c}{} \\ 
   13531 &  0.118$\pm$0.036 &  0.121$\pm$0.024 &  0.229$\pm$0.017 & \multicolumn{9}{c}{} \\ 
   13931 & -0.008$\pm$0.022 &  0.013$\pm$0.021 &  0.046$\pm$0.025 & \multicolumn{9}{c}{} \\ 
   32963 &  0.090$\pm$0.020 &  0.064$\pm$0.018 &  0.071$\pm$0.017 & \multicolumn{9}{c}{} \\ 
   33636 & -0.027$\pm$0.028 & -0.044$\pm$0.025 &  0.085$\pm$0.020 & \multicolumn{9}{c}{} \\ 
   43162 &  0.216$\pm$0.041 &  0.180$\pm$0.029 &  0.313$\pm$0.026 & \multicolumn{9}{c}{} \\ 
   45184 &  0.077$\pm$0.028 &  0.092$\pm$0.029 &  0.137$\pm$0.023 & \multicolumn{9}{c}{} \\ 
   87359 &  0.190$\pm$0.016 &  0.147$\pm$0.025 &  0.150$\pm$0.019 & \multicolumn{9}{c}{} \\ 
   95128 & -0.013$\pm$0.015 & -0.022$\pm$0.014 &  0.007$\pm$0.016 & \multicolumn{9}{c}{} \\ 
  98618 &  0.063$\pm$0.020 &  0.059$\pm$0.016 &  0.074$\pm$0.016 & \multicolumn{9}{c}{} \\ 
   106252 & -0.129$\pm$0.017 & -0.102$\pm$0.017 & -0.082$\pm$0.017 & \multicolumn{9}{c}{} \\ 
   112257* & -0.097$\pm$0.016 & -0.036$\pm$0.018 & -0.001$\pm$0.017 & \multicolumn{9}{c}{} \\ 
   140538 &  0.152$\pm$0.028 &  0.151$\pm$0.033 &  0.016$\pm$0.025 & \multicolumn{9}{c}{} \\ 
   143436 &  0.105$\pm$0.019 &  0.072$\pm$0.027 &  0.119$\pm$0.024 & \multicolumn{9}{c}{} \\ \cline{1-4}
\end{tabular}
\begin{tablenotes}
      \tiny
      \item Note: (*) high-$\alpha$ metal-rich star.     
      \end{tablenotes}
    \end{threeparttable}
\end{sidewaystable*}

\begin{sidewaystable*}
\begin{threeparttable}
\caption{ [X/Fe] abundance ratios}\label{XFe_ratios}
        \centering 
\tiny
\begin{tabular}{l|cccccccccccc} 
\hline\hline             
HD & C & CH & $\widehat{\textrm{C}}$ & O & Na & Mg & Al & Si & S & K & Ca & Sc II  \\ \hline
   9986 & -0.072$\pm$0.015 & -0.055$\pm$0.012 & -0.065$\pm$0.009 & -0.059$\pm$0.015 & -0.038$\pm$0.015 & -0.016$\pm$0.009 & -0.005$\pm$0.006 & -0.007$\pm$0.007 & -0.009$\pm$0.025 & -0.018$\pm$0.017 &  0.005$\pm$0.007 &  0.000$\pm$0.017 \\
   13531 & -0.050$\pm$0.035 & -0.060$\pm$0.046 & -0.053$\pm$0.028 & -0.042$\pm$0.037 & -0.095$\pm$0.012 & -0.036$\pm$0.007 & -0.044$\pm$0.018 & -0.038$\pm$0.014 & -0.019$\pm$0.028 & -0.012$\pm$0.016 &  0.026$\pm$0.010 & -0.085$\pm$0.026 \\
   13931 & -0.009$\pm$0.026 &  0.001$\pm$0.021 & -0.004$\pm$0.016 & -0.024$\pm$0.023 & -0.001$\pm$0.016 &  0.029$\pm$0.012 &  0.034$\pm$0.017 &  0.015$\pm$0.010 & -0.027$\pm$0.027 &  0.071$\pm$0.011 &  0.015$\pm$0.008 &  0.056$\pm$0.019 \\
   32963 & -0.005$\pm$0.021 & -0.021$\pm$0.018 & -0.013$\pm$0.014 & -0.023$\pm$0.015 &  0.001$\pm$0.014 &  0.007$\pm$0.010 &  0.041$\pm$0.020 &  0.011$\pm$0.009 &  0.018$\pm$0.022 &  0.002$\pm$0.015 &  0.004$\pm$0.005 &  0.035$\pm$0.014 \\
   33636 & -0.077$\pm$0.037 & -0.006$\pm$0.063 & -0.068$\pm$0.032 & -0.052$\pm$0.040 & -0.066$\pm$0.015 & -0.028$\pm$0.013 & -0.016$\pm$0.010 &  0.006$\pm$0.014 & -0.104$\pm$0.043 &  0.084$\pm$0.021 &  0.034$\pm$0.008 & -0.002$\pm$0.029 \\
   43162 & -0.238$\pm$0.048 & -0.199$\pm$0.029 & -0.221$\pm$0.025 & -0.067$\pm$0.054 & -0.102$\pm$0.026 & -0.041$\pm$0.015 & -0.058$\pm$0.027 & -0.036$\pm$0.023 & -0.077$\pm$0.052 &  0.020$\pm$0.015 &  0.040$\pm$0.010 & -0.063$\pm$0.034 \\
   45184 & -0.108$\pm$0.035 & -0.061$\pm$0.016 & -0.082$\pm$0.015 & -0.092$\pm$0.038 & -0.041$\pm$0.012 & -0.025$\pm$0.014 & -0.021$\pm$0.010 &  0.001$\pm$0.015 & -0.049$\pm$0.045 &  0.025$\pm$0.021 &  0.010$\pm$0.008 & -0.007$\pm$0.030 \\
   87359 & -0.098$\pm$0.029 & -0.094$\pm$0.029 & -0.097$\pm$0.021 & -0.088$\pm$0.026 & -0.060$\pm$0.007 & -0.006$\pm$0.012 &  0.014$\pm$0.007 & -0.007$\pm$0.011 & -0.004$\pm$0.025 & -0.025$\pm$0.019 &  0.013$\pm$0.010 & -0.002$\pm$0.030 \\
   95128 &  0.005$\pm$0.020 &  0.023$\pm$0.013 &  0.016$\pm$0.011 &  0.018$\pm$0.016 &  0.027$\pm$0.005 & -0.006$\pm$0.009 &  0.020$\pm$0.014 &  0.006$\pm$0.007 & -0.024$\pm$0.016 &  0.065$\pm$0.012 & -0.003$\pm$0.008 &  0.039$\pm$0.016 \\
   98618 & -0.072$\pm$0.021 & -0.043$\pm$0.018 & -0.061$\pm$0.014 & -0.048$\pm$0.020 & -0.038$\pm$0.011 & -0.006$\pm$0.010 &  0.028$\pm$0.008 & -0.016$\pm$0.009 & -0.032$\pm$0.028 & -0.009$\pm$0.010 &  0.004$\pm$0.005 &  0.021$\pm$0.017 \\
   106252 &  0.028$\pm$0.016 &  0.023$\pm$0.017 &  0.027$\pm$0.012 &  0.059$\pm$0.020 &  0.029$\pm$0.013 & -0.008$\pm$0.007 &  0.015$\pm$0.005 &  0.018$\pm$0.009 &  0.062$\pm$0.027 &  0.072$\pm$0.017 & -0.003$\pm$0.009 &  0.050$\pm$0.014 \\
   112257* &  0.070$\pm$0.027 &  0.014$\pm$0.021 &  0.040$\pm$0.017 &  0.112$\pm$0.021 &  0.007$\pm$0.011 &  0.110$\pm$0.013 &  0.128$\pm$0.017 &  0.040$\pm$0.010 &  0.123$\pm$0.022 &  0.051$\pm$0.011 &  0.055$\pm$0.005 &  0.101$\pm$0.014 \\ 
   140538 & -0.149$\pm$0.037 & -0.104$\pm$0.015 & -0.119$\pm$0.014 & -0.098$\pm$0.033 & -0.066$\pm$0.011 & -0.012$\pm$0.014 &  0.004$\pm$0.010 & -0.018$\pm$0.015 & -0.082$\pm$0.040 & -0.052$\pm$0.020 &  0.016$\pm$0.009 & -0.011$\pm$0.030 \\
   143436 & -0.055$\pm$0.033 & -0.015$\pm$0.017 & -0.029$\pm$0.015 & -0.067$\pm$0.023 & -0.055$\pm$0.016 & -0.041$\pm$0.012 & -0.011$\pm$0.008 & -0.014$\pm$0.013 & -0.022$\pm$0.034 &  0.008$\pm$0.024 & -0.002$\pm$0.009 & -0.032$\pm$0.032 \\ \hline \hline
HD & Ti I &  Ti II & $\widehat{\textrm{Ti}}$ & V & Cr I & Cr II & $\widehat{\textrm{Cr}}$ & Mn & Co & Ni  & Cu & Zn \\ \hline
   9986 & -0.004$\pm$0.006 & -0.004$\pm$0.018 & -0.004$\pm$0.006 & -0.014$\pm$0.007 &  0.011$\pm$0.005 & -0.007$\pm$0.018 &  0.007$\pm$0.005 & -0.009$\pm$0.006 & -0.009$\pm$0.010 & -0.019$\pm$0.004 & -0.045$\pm$0.014 & -0.024$\pm$0.008 \\
   13531 &  0.005$\pm$0.009 & -0.039$\pm$0.022 & -0.001$\pm$0.008 &  0.013$\pm$0.012 &  0.026$\pm$0.012 & -0.001$\pm$0.026 &  0.015$\pm$0.011 & -0.049$\pm$0.009 & -0.072$\pm$0.013 & -0.066$\pm$0.010 & -0.093$\pm$0.011 & -0.098$\pm$0.025 \\
   13931 &  0.012$\pm$0.006 &  0.028$\pm$0.018 &  0.017$\pm$0.006 &  0.020$\pm$0.030 &  0.011$\pm$0.011 & -0.007$\pm$0.020 &  0.004$\pm$0.010 & -0.017$\pm$0.006 &  0.020$\pm$0.008 &  0.019$\pm$0.006 &  0.022$\pm$0.005 &  0.003$\pm$0.026 \\
   32963 &  0.020$\pm$0.005 &  0.012$\pm$0.015 &  0.018$\pm$0.005 &  0.010$\pm$0.008 & -0.003$\pm$0.007 & -0.009$\pm$0.016 & -0.005$\pm$0.006 &  0.014$\pm$0.009 &  0.026$\pm$0.009 &  0.016$\pm$0.006 &  0.027$\pm$0.006 &  0.025$\pm$0.012 \\
   33636 &  0.029$\pm$0.009 &  0.011$\pm$0.029 &  0.022$\pm$0.009 & -0.002$\pm$0.015 & -0.008$\pm$0.010 & -0.012$\pm$0.036 & -0.009$\pm$0.010 & -0.096$\pm$0.010 & -0.026$\pm$0.011 & -0.046$\pm$0.008 & -0.117$\pm$0.037 & -0.088$\pm$0.040 \\
   43162 &  0.001$\pm$0.012 & -0.040$\pm$0.030 & -0.026$\pm$0.011 & -0.010$\pm$0.017 &  0.016$\pm$0.012 & -0.012$\pm$0.035 &  0.000$\pm$0.011 & -0.062$\pm$0.014 & -0.098$\pm$0.012 & -0.077$\pm$0.013 & -0.130$\pm$0.022 & -0.076$\pm$0.032 \\
   45184 & -0.003$\pm$0.007 & -0.005$\pm$0.030 & -0.004$\pm$0.007 & -0.002$\pm$0.013 &  0.000$\pm$0.005 &  0.000$\pm$0.032 &  0.000$\pm$0.005 & -0.035$\pm$0.007 & -0.001$\pm$0.019 & -0.005$\pm$0.008 & -0.055$\pm$0.012 & -0.025$\pm$0.016 \\
   87359 &  0.046$\pm$0.007 &  0.014$\pm$0.028 &  0.040$\pm$0.007 &  0.023$\pm$0.010 &  0.010$\pm$0.007 & -0.006$\pm$0.029 &  0.007$\pm$0.007 & -0.007$\pm$0.008 & -0.020$\pm$0.012 & -0.026$\pm$0.007 & -0.021$\pm$0.012 & -0.055$\pm$0.014 \\
   95128 & -0.015$\pm$0.005 &  0.014$\pm$0.014 & -0.007$\pm$0.005 &  0.019$\pm$0.017 & -0.006$\pm$0.006 & -0.005$\pm$0.017 & -0.006$\pm$0.006 &  0.012$\pm$0.014 &  0.009$\pm$0.009 &  0.005$\pm$0.004 &  0.015$\pm$0.011 & -0.005$\pm$0.015 \\
   98618 &  0.009$\pm$0.006 &  0.022$\pm$0.017 &  0.013$\pm$0.006 &  0.016$\pm$0.009 & -0.015$\pm$0.006 &  0.002$\pm$0.019 & -0.010$\pm$0.006 & -0.032$\pm$0.007 &  0.002$\pm$0.007 &  0.002$\pm$0.005 & -0.025$\pm$0.013 & -0.041$\pm$0.026 \\
   106252 &  0.007$\pm$0.008 &  0.032$\pm$0.016 &  0.015$\pm$0.007 &  0.014$\pm$0.009 & -0.003$\pm$0.011 & -0.002$\pm$0.030 & -0.003$\pm$0.010 & -0.041$\pm$0.006 &  0.039$\pm$0.012 &  0.004$\pm$0.006 & -0.020$\pm$0.014 & -0.007$\pm$0.007 \\
   112257* &  0.101$\pm$0.008 &  0.100$\pm$0.016 &  0.101$\pm$0.007 &  0.043$\pm$0.008 &  0.022$\pm$0.005 & -0.017$\pm$0.019 &  0.012$\pm$0.005 & -0.048$\pm$0.012 &  0.047$\pm$0.009 &  0.014$\pm$0.008 &  0.054$\pm$0.017 &  0.111$\pm$0.010 \\ 
   140538 &  0.036$\pm$0.007 &  0.006$\pm$0.029 &  0.028$\pm$0.007 &  0.017$\pm$0.008 &  0.021$\pm$0.007 &  0.007$\pm$0.032 &  0.018$\pm$0.007 & -0.010$\pm$0.008 & -0.021$\pm$0.009 & -0.031$\pm$0.007 & -0.065$\pm$0.014 & -0.071$\pm$0.020 \\
   143436 & -0.016$\pm$0.008 & -0.020$\pm$0.030 & -0.017$\pm$0.008 & -0.015$\pm$0.011 &  0.001$\pm$0.007 & -0.013$\pm$0.031 & -0.002$\pm$0.007 &  0.004$\pm$0.013 & -0.038$\pm$0.013 & -0.025$\pm$0.007 & -0.033$\pm$0.017 & -0.048$\pm$0.018 \\ \hline \hline

\cline{1-4} 

\cline{1-4}
HD & Sr & Y II & Ba \\ \cline{1-4}
   9986 &  0.072$\pm$0.012 &  0.057$\pm$0.018 &  0.061$\pm$0.010 & \multicolumn{9}{c}{} \\ 
   13531 &  0.098$\pm$0.032 &  0.101$\pm$0.025 &  0.209$\pm$0.016 & \multicolumn{9}{c}{} \\ 
   13931 & -0.075$\pm$0.019 & -0.054$\pm$0.020 & -0.021$\pm$0.023 & \multicolumn{9}{c}{} \\ 
   32963 &  0.002$\pm$0.016 & -0.024$\pm$0.015 & -0.017$\pm$0.012 & \multicolumn{9}{c}{} \\ 
   33636 &  0.055$\pm$0.021 &  0.038$\pm$0.028 &  0.167$\pm$0.019 & \multicolumn{9}{c}{} \\ 
   43162 &  0.159$\pm$0.024 &  0.123$\pm$0.030 &  0.256$\pm$0.022 & \multicolumn{9}{c}{} \\ 
   45184 &  0.007$\pm$0.019 &  0.022$\pm$0.031 &  0.067$\pm$0.020 & \multicolumn{9}{c}{} \\ 
   87359 &  0.125$\pm$0.011 &  0.082$\pm$0.028 &  0.085$\pm$0.019 & \multicolumn{9}{c}{} \\ 
   95128 & -0.035$\pm$0.014 & -0.044$\pm$0.016 & -0.015$\pm$0.016 & \multicolumn{9}{c}{} \\ 
   98618 &  0.009$\pm$0.017 &  0.005$\pm$0.017 &  0.020$\pm$0.015 & \multicolumn{9}{c}{} \\ 
   106252 & -0.060$\pm$0.014 & -0.033$\pm$0.017 & -0.013$\pm$0.015 & \multicolumn{9}{c}{} \\ 
   112257* & -0.094$\pm$0.012 & -0.033$\pm$0.019 &  0.002$\pm$0.016 & \multicolumn{9}{c}{} \\ 
   140538 &  0.093$\pm$0.018 &  0.092$\pm$0.030 & -0.043$\pm$0.019 & \multicolumn{9}{c}{} \\ 
   143436 &  0.061$\pm$0.010 &  0.028$\pm$0.028 &  0.075$\pm$0.022 & \multicolumn{9}{c}{} \\ \cline{1-4} 
\end{tabular}
\begin{tablenotes}
      \tiny
      \item Note: (*) high-$\alpha$ metal-rich star.     
      \end{tablenotes}
    \end{threeparttable}
\end{sidewaystable*}

\end{document}